\documentclass[preprint,authoryear,12pt]{elsarticle}

\usepackage{amssymb}
\usepackage{color}

\journal{Physics of the Dark Universe}

\begin{document}

\begin{frontmatter}



\title{Cold dark matter halos in Multi-coupled Dark Energy cosmologies: structural and statistical properties}


\author{Marco Baldi}
\address{Dipartimento di Fisica e Astronomia, Universit\`a di Bologna,
viale Berti Pichat 6/2 -- I-40126, Bologna, Italy;\\
INAF - Osservatorio Astronomico di Bologna, via Ranzani 1, I-40127 Bologna, Italy;\\
INFN - Sezione di Bologna, viale Berti Pichat 6/2, I-40127 Bologna, Italy.}
\begin{abstract}
The recently proposed Multi-coupled Dark Energy (McDE) scenario -- characterised by two distinct Cold Dark Matter (CDM) particle species with opposite couplings to a Dark Energy scalar field -- introduces a number of novel features in the small-scale dynamics of cosmic structures, most noticeably the simultaneous existence of both {\em attractive} and {\em repulsive} fifth-forces. Such small-scale features are expected to imprint possibly observable footprints on nonlinear cosmic structures, that might provide a direct way to test the scenario. In order to unveil such footprints, we have performed the first suite of high-resolution N-body simulations of McDE cosmologies, covering the coupling range $|\beta |\leq 1$. We find that for coupling values corresponding to fifth-forces weaker than standard gravity, the impact on structure formation is  very mild, thereby showing a new type of screening mechanism for long-range scalar interactions. On the contrary, for fifth-forces comparable to or stronger than standard gravity a number of effects appear in the statistical and structural properties of CDM halos. Collapsed structures start to fragment into pairs of smaller objects that move on different trajectories, providing a direct evidence of the violation of the weak equivalence principle. Consequently, the relative abundance of halos of different masses is significantly modified. For sufficiently large coupling values, the expected number of clusters is strongly suppressed, which might alleviate the present tension between CMB- and cluster-based cosmological constraints. Finally, the internal structure of halos is also modified, with a significant suppression of the inner overdensity, and a progressive segregation of the two CDM species.
\end{abstract}


\end{frontmatter}


\section{Introduction}
\label{sec:intro}

Any extension of the standard cosmological model that aims to overcome the fine-tuning problems
of the cosmological constant must introduce additional degrees of freedom playing the role of Dark Energy (DE)
in the form of a new dynamical field with defined clustering and interaction properties \citep[][]{Euclid_TWG}
or as a low-energy modification of the laws of gravity. 
Despite the ever increasing accuracy of observational data of different kinds \citep[see e.g.][]{Vikhlinin_etal_2009b,Reid_etal_2010,Dunkley_etal_2010,wmap9,Planck_016}
showing full consistency with the expected
behaviour of a $\Lambda $CDM cosmology, such extended scenarios retain their appeal as the only possible 
alternative to anthropic arguments in explaining the observed value of the Dark Energy density at present.
Nonetheless, large deviations from the standard background and linear perturbations evolution have been
progressively ruled out by data, thereby significantly restricting the parameter space of a wide range of extended
cosmologies to the level where potentially observable features would be ever hardly detectable. In this respect, a considerable
interest has been attracted in the last years by extended cosmologies characterised by various types of screening mechanisms capable of
recovering the standard $\Lambda $CDM behaviour in the appropriate regimes that are tightly constrained by observations, still allowing
for possibly detectable deviations elsewhere. 

For modified theories of gravity (including e.g. $f(R)$ models)  the recovery of the standard behaviour can be enforced
both at cosmological scales, by selecting the model's parameters in order to reproduce as close as possible the standard $\Lambda $CDM
expansion history \citep[see e.g.][]{Hu_Sawicki_2007}, and within overdense regions of the universe (like our Galaxy), through various types of screening mechanisms \citep[as the {\em chameleon}, {\em symmetron}, and {\em Vainshtein} mechanisms, see e.g.][respectively]{Khoury_Weltman_2004,Hinterbichler_Khoury_2010,Deffayet_etal_2002}. Such construction allows compatibility with
both cosmological observations and solar-system tests of gravity \citep[see e.g.][]{Bertotti_Iess_Tortora_2003,Will_2005,Amendola_Tsujikawa_2008,Capozziello_Tsujikawa_2008}.

Alternatively, if the new degree of freedom
associated with DE has a selective interaction to Dark Matter, leaving baryonic particles uncoupled \citep[][]{Damour_Gibbons_Gundlach_1990},
as for the case of coupled Dark Energy (cDE) models \citep[][]{Wetterich_1995,Amendola_2000,Farrar2004}, solar system tests of gravity do not apply, and the model is mainly constrained by its impact on cosmological observables \citep[][]{Bean_etal_2008,Xia_2009,Baldi_Viel_2010,Pettorino_etal_2012}. However, even in this case such constraints are capable to bind the DE-CDM interaction to a level that
makes its predicted observational signatures at small scales hardly detectable with the presently available observational precision, at least for the most widely considered case of a constant coupling. 

Several simple extensions of such standard
cDE scenario have been proposed in recent years, with the aim to explore the range of plausible non-standard DE cosmologies that might allow for testable predictions at the scales of structure formation
without conflicting with the ever more stringent bounds arising from cosmological probes. These extensions include the possibility of time-dependent coupling functions \citep[see e.g.][]{Amendola_2004,Baldi_2011a}, resulting
in a strongly suppressed impact of the interaction on the background cosmic evolution even in the presence of significant effects on the formation and evolution of nonlinear structures; alternatively, the possibility of having
multiple species of CDM particles interacting with DE through individual coupling functions have also been proposed \citep[see e.g.][]{Brookfield_VanDeBruck_Hall_2008}. The latter scenarios are characterised by attractor solutions that suppress the effective coupling
to DE during matter domination, thereby making the model consistent with the observed expansion history \citep[][]{Piloyan_etal_2013}. 

Although such ``extended" cDE models might appear in general less appealing than
standard quintessence and cDE scenarios -- as they generally involve one or more additional free parameters 
associated with the specific time evolution of the coupling function or to the individual couplings of different CDM species --
a particular realisation of the latter class of models requiring no additional parameters as compared to standard cDE cosmologies has been recently proposed \citep[][]{Baldi_2012a}, 
and termed the ``Multi-coupled DE" (McDE, hereafter) scenario. This features only two distinct CDM particle species with opposite constant couplings to a DE scalar field, and has been shown
to provide a very effective screening of the interaction during matter domination even for very large values of the coupling constant $\beta $ \citep[see again][]{Piloyan_etal_2013}. Such screening is then broken at the onset of DE domination, thereby
providing a time-dependent effective coupling without imposing {\em a priori} any time evolution of the coupling function. 
{It is important to stress again here that the idea of a non-universal coupling between a cosmic scalar and distinct matter fluids is not a new concept that characterises only the McDE scenario: indeed, such idea is at the very basis of the whole class of Coupled Quintessence models, where a non-universal coupling has to be invoked \citep[see again][]{Damour_Gibbons_Gundlach_1990} in order to keep baryonic particles minimally-coupled so to evade constraints on scalar-tensor theories from local tests of gravity.
The truly distinctive feature of the McDE scenario, however,} is that the presence of {coupling
constants with an opposite sign} for the two different CDM particle species determines the existence of both attractive and {\em repulsive} fifth-forces between CDM particles, differently from all standard cDE models where 
fifth-forces are always attractive. 

{It is also worth mentioning how the possibility of a multi-particle nature of the CDM cosmic fraction has been proposed in several different theoretical contexts \citep[see e.g. the recent work by][]{Chialva_Dev_Mazumdar_2013}, considering matter species that differ from each other in various physical properties. For instance, the possibility of multi-particle dark matter models featuring a cold and a hot component has been proposed \citep[see e.g.][]{Anderhalden_etal_2012,Maccio_etal_2013} as a possible solution of the small-scale problems of the $\Lambda $CDM scenario, which in some specific realisations include also a non-vanishing coupling of the cold species to a DE scalar field \citep[as for the model proposed by][]{Bonometto_Sassi_LaVacca_2012,Bonometto_Mainini_2014}. At a more fundamental level, models with multiple CDM species with different couplings to a cosmic light scalar might arise in the context of string-inspired cosmological scenarios \citep[see e.g.][]{Brandenberger_Vafa_1989} as proposed by \cite{Gubser_Peebles_2004} and subsequently extensively discussed by  \cite{Gubser2004,Farrar2004,Nusser_Gubser_Peebles_2005}. The McDE scenario, in particular, represents the simplest realisation of the model proposed by \cite{Gubser_Peebles_2004} which has as its most characteristic footprint the existence of long-range attractive and repulsive fifth-forces.}

{For coupling values that appear to be consistent with the observed background and linear perturbations evolution \citep[see e.g. the recent work by][]{Piloyan_etal_2014}, such long-range fifth-forces might have a strength comparable to standard gravity, giving rise to a very peculiar phenomenology at the level of linear and nonlinear structure formation \citep[][]{Baldi_2013}. 
The work by \cite{Piloyan_etal_2014}, however, has also shown how the evolution of linear density perturbations suddenly deviates from the standard $\Lambda $CDM behaviour when the coupling grows beyond the value corresponding to a gravitational strength of the associated fifth-forces, thereby allowing to place much tighter constraints on the coupling itself through measurements of the linear growth rate as compared to the extremely loose bounds derived from the background expansion history alone.}
This phenomenology
has been already explored with both analytical and numerical tools, and in particular the nonlinear evolution of large-scale structures has been investigated with some first low-resolution N-body simulations 
\citep[][]{Baldi_2013} aimed at coarsely sampling the model's parameter space and highlighting its most prominent effects on the large-scale shape of the cosmic web. Such analysis has shown that
fifth-forces with the same strength as standard gravity appear to still have a relatively mild impact on the overall distribution of nonlinear structures in McDE cosmologies, and allowed to observe for the first time the 
halo fragmentation process associated with the repulsive interaction between the two different CDM particle species. The details of such small scale effects, however, could not be observed due to the limited resolution of those early N-body simulations, and their proper investigation demands higher resolution runs for the range of parameters that already showed to provide a reasonable behaviour of structure formation at large scales.

With the present work, we move in such direction by presenting the first high-resolution N-body simulations of McDE models ever performed, and investigating 
the effects of such cosmologies on the statistical and structural properties of CDM halos for several different values of the coupling constant $\beta $. \\

The paper is organised as follows. In Section~\ref{sec:McDE} we review the main features of McDE models at the background and linear perturbations level; in Section~\ref{sec:sims} we describe the numerical setup of our high-resolution N-body simulations, and illustrate the basic post-processing analysis performed on the simulations outputs; in Section~\ref{sec:results} we discuss the results of our investigation on a number of potentially observable statistical and structural properties of CDM halos. Finally, in Section~\ref{sec:concl} we drive our conclusions.

\section{Multi-coupled Dark Energy cosmologies}
\label{sec:McDE}

Here we present a short overview on the basic equations of McDE cosmologies at the background and linear perturbations level,
with the main aim to provide a quick reference and to set the notation that will be used throughout the paper. A comprehensive 
discussion of the background and linear perturbations features of McDE goes beyond the scope of the present paper and
 has been presented in two previous publications \citep[][]{Baldi_2012a,Piloyan_etal_2013} to which we refer the interested reader for a thorough introduction.\\
 
 We consider a series of flat cosmologies where the role of DE is played by a classical scalar field $\phi $ moving in a self-interaction
 potential $V(\phi )$. Without loss of generality, we will restrict to the case of an exponential \citep[][]{Lucchin_Matarrese_1984,Ferreira_Joyce_1998}
 potential of the form:
 \begin{equation}
V(\phi ) = Ae^{-\alpha \phi /M_{\rm Pl}}\,. 
\end{equation}
Additionally, our cosmological models include radiation and two families of CDM particles, characterised by opposite couplings to the DE scalar field.
For simplicity, we discard the baryonic component of the universe which is not expected to affect our results at least at the level of accuracy reached within the present analysis. 
{This expectation has been recently confirmed by \cite{Piloyan_etal_2014}, where the full system of both background and linear perturbations equations including baryons has been investigated in detail, showing how the inclusion of the uncoupled baryonic component has a very minor effect on the main characteristic features of the model predicted by a simplified CDM-only system.}

With such assumptions, the background evolution of the system is governed by the following set of dynamic equations:
\begin{eqnarray}
\label{klein_gordon}
\ddot{\phi } + 3H\dot{\phi } + \frac{dV}{d\phi } &=& +C \rho _{+} - C \rho _{-}\,, \\
\label{continuity_plus}
\dot{\rho }_{+} + 3H\rho _{+} &=& -C \dot{\phi }\rho _{+} \,, \\
\label{continuity_minus}
\dot{\rho }_{-} + 3H\rho _{-} &=& +C \dot{\phi }\rho _{-} \,, \\
\label{continuity_radiation}
\dot{\rho }_{r} + 4H\rho _{r} &=& 0\,, \\
\label{friedmann}
3H^{2} &=& \frac{1}{M_{{\rm Pl}}^{2}}\left( \rho _{r} + \rho _{+} + \rho _{-} + \rho _{\phi} \right)\,,
\end{eqnarray}
where an overdot indicates a derivative with respect to the cosmic time $t$, $H$ is the Hubble function, the ``$+$" and ``$-$" subscripts denote
the two different CDM species according to the sign of the coupling to the DE scalar field, and C is the coupling strength, which we will
indicate in the rest of the paper in its dimensionless form denoted by the parameter $\beta $:
\begin{equation}
\beta \equiv \sqrt{\frac{3}{2}}M_{\rm Pl}C \,,
\end{equation}
with $M_{\rm Pl} \equiv 1/\sqrt{8\pi G}$ being the reduced Planck mass and  $G$ the Newton's constant.
The main peculiarity of such system is provided by the existence -- during matter domination -- of an attractor solution characterised by an equal abundance
of the two different CDM particle species \citep[see][for a detailed analysis of the phase space of McDE models]{Baldi_2012a,Piloyan_etal_2013}, such that the effective coupling is strongly 
suppressed and the cosmic background evolution is indistinguishable from that of an uncoupled system.
Then, at the onset of DE domination the system leaves the attractor solution and progressively develops an asymmetry between the two CDM species, thereby giving rise
to a late-time non-vanishing effective coupling with an increasing strength that might affect structure formation processes at low redshifts still leaving almost completely unaffected the background expansion history, even for very large values of the nominal coupling $\beta $ \citep[see again][]{Baldi_2012a,Piloyan_etal_2013}.\\

At the level of sub-horizon linear density perturbations, the opposite coupling constants of the two different CDM particle types introduce a few correction terms to the evolution equations for the density contrast,
that read:
\begin{eqnarray}
\label{gf_plus}
\ddot{\delta }_{+} = -2H\left[ 1 - \beta \frac{\dot{\phi }}{H\sqrt{6}}\right] \dot{\delta }_{+} + 4\pi G \left[ \rho _{-}\delta _{-} \Gamma_{R} + \rho _{+}\delta _{+}\Gamma_{A}\right] \,, \\
\label{gf_minus}
\ddot{\delta }_{-} = -2H\left[ 1 + \beta \frac{\dot{\phi }}{H\sqrt{6}}\right] \dot{\delta }_{-} + 4\pi G \left[ \rho _{-}\delta _{-} \Gamma _{A} + \rho _{+}\delta _{+}\Gamma_{R}\right]\,.
\end{eqnarray}
where the $\Gamma $ factors are defined as:
\begin{equation}
\label{def_gamma}
\Gamma _{A} \equiv 1 + \frac{4}{3}\beta ^{2}\,, \quad \Gamma _{R}\equiv 1 - \frac{4}{3}\beta ^{2} \,,
\end{equation}
and represent attractive ($\Gamma _{A}$) or repulsive ($\Gamma _{R}$) corrections to gravity due to the
long-range fifth-force mediated by the DE scalar field. It is interesting to notice that along the matter-dominated attractor -- where the condition $\rho _{+} = \rho _{-}$ holds -- and for 
adiabatic initial conditions (i.e. $\rho _{+}\delta _{+} = \rho _{-}\delta _{-}$) the fifth-force corrections exactly vanish in both equations, so that the gravitational source term for the evolution of CDM density perturbations
is simply given by the standard Newtonian gravitational potential. In this respect, the system appears to be symmetric in the two CDM species and would remain effectively uncoupled as long as the universe is matter dominated.
However, the second terms in the first square brackets on the right-hand side of Eqs.~\ref{gf_plus}-\ref{gf_minus} clearly break the symmetry of the two equations and will consequently move the system away from the condition of adiabiaticity whenever the DE scalar field evolves in time -- even along the attractor solution -- thereby triggering also the fifth-force corrections that do no longer vanish for the case $\rho _{+}\delta _{+}\neq \rho_{-}\delta_{-}$. 

The screening of the effective coupling that is provided by the background attractor solution is therefore broken at the level of linear perturbations, allowing to distinguish (at least in principle) a McDE cosmology from
an uncoupled system \citep[as first highlighted by][]{Brookfield_VanDeBruck_Hall_2008}. Nonetheless, since the scalar filed moves very slowly along the matter dominated attractor, the deviation from adiabaticity during matter domination remains small unless the repulsive corrections in Eqs.~\ref{gf_plus}-\ref{gf_minus} are large enough to overcome the attractive pull of standard gravity. Such condition corresponds to coupling values $|\beta | > \sqrt{3}/2$, and in \citet{Baldi_2012a} it was shown that couplings as large as $|\beta |=1$ still do not induce any appreciable deviation of the total matter growth factor $(\Omega _{+}\delta_{+} + \Omega _{-}\delta _{-})/(\Omega _{+} + \Omega _{-})$ from the uncoupled solution down to $z=0$, while further increasing the coupling value leads to a very rapid increase of the total growth factor at low redshifts with e.g. $|\beta |=1.5$ resulting in a 13\% enhancement of the total growth factor at $z=0$ and $|\beta |=2$ already determining an enhancement of a factor $60$.

The coupling range $|\beta |\leq 1$ of McDE models therefore appears to be very effectively screened both at the background and linear perturbations level, with no significative deviations in the expansion history and in the linear growth rate from the case of an uncoupled system, which in turn can be tuned to be indistinguishable from a standard $\Lambda $CDM cosmology. Such range of couplings has then been explored
also through a first suite of low-resolution N-body simulations \citep[][]{Baldi_2013}, with the aim to find possible characteristic signatures of the model at nonlinear scales and further constrain its parameter space.
Indeed, such first numerical investigation was capable to capture some basic features of McDE scenarios as e.g. the low-redshift fragmentation of collapsed structures into smaller objects driven by the repulsive fifth-force corrections for sufficiently large coupling values; additionally, it was shown that a significant suppression of power appears at small scales ($k \gtrsim 1\, h/$Mpc) for coupling values above the gravitational coupling threshold
of $|\beta |=\sqrt{3}/2$. This value defines the border between couplings for which fifth-forces are weaker ($|\beta |<\sqrt{3}/2$) and stronger ($|\beta |>\sqrt{3}/2$) than standard gravity, and corresponds to the very peculiar situation where repulsive corrections ($\Gamma _{\rm R}$) exactly balance the gravitational attraction such that opposite species of CDM particles do not exert any long-range force on each other. As we will see below, such turning point in the McDE parameter space gives rise to a very interesting and peculiar phenomenology at the level of the formation and dynamical evolution of nonlinear cosmic structures at small scales.\\

In the present paper, we aim to explore the coupling range $|\beta |\leq 1$ of McDE cosmologies down to the highly nonlinear regime of cosmic structure formation to a much finer level of detail than allowed by the first
low-resolution simulations of \citet{Baldi_2013}, with a particular focus on the dynamical evolution of small-scale structures and on the statistical and structural properties of CDM halos forming in McDE cosmologies for values of $|\beta |$ around the gravitational coupling threshold $|\beta |=\sqrt{3}/2$. To this end, we have run a series of high-resolution N-body simulations that are described in detail in the next Section.

\section{The simulations}
\label{sec:sims}
\begin{table}
\begin{center}
\begin{tabular}{cc}
\hline
Parameter & Value\\
\hline
$H_{0}$ & 70.3 km s$^{-1}$ Mpc$^{-1}$\\
$\Omega _{\rm CDM} $ & 0.226 \\
$\Omega _{\rm DE} $ & 0.729 \\
${\cal A}_{s}$ & $2.42 \times 10^{-9}$\\
$ \Omega _{b} $ & 0.0451 \\
$n_{s}$ & 0.966\\
\hline
\end{tabular}
\end{center}
\caption{The set of cosmological parameters at $z=0$ assumed for all the models investigated in the present work, consistent with the 
WMAP7 Maximum Likelihood results \citep[][]{wmap7} based on CMB data alone.}
\label{tab:parameters}
\end{table}
Following the approach described in \citet{Baldi_2013}, we have made use of a modified version of the widely used parallel Tree-PM and SPH N-body code
{\small GADGET-3} \citep[][]{gadget} to run a series of 7 high-resolution N-body simulations of structure formation in a periodic cosmological box of $100$ Mpc$/h$ aside, filled with $512^{3}$ particles for each CDM species, for a total of 
$\approx 2.7\times 10^{8}$ particles. The mass resolution at the starting redshift ($z_{\rm i}=99$) is $m_{\pm } = 2.8\times 10^{8}$ M$_{\odot}/h$ and
the gravitational softening has been set to $\epsilon _{\rm g} = 5$ kpc$/h$ corresponding to about $0.05$ times the mean inter-particle separation. The adopted N-body code allows to follow all the characteristic effects of a wide range of DE scenarios,
including the McDE model under investigation, and has been widely employed in the last years to run cosmological and hydrodynamical simulations of a variety of interacting DE cosmologies \citep[see e.g.][]{Baldi_etal_2010,Baldi_2011a,Baldi_2011c,Baldi_etal_2011a,Baldi_Lee_Maccio_2011,Baldi_Pettorino_2011,Baldi_Viel_2010,CoDECS,Lee_Baldi_2011} and has therefore been very extensively tested. The McDE models included in our suite cover the 
$|\beta | \leq 1$ range of couplings with a non-uniform spacing that more finely samples the coupling values just below the gravitational coupling barrier $|\beta |=\sqrt{3}/2$. More specifically, we have run simulations for the coupling values
$|\beta |=\{ 0\,, 1/2\,, 7/10\,, 3/4\,, 8/10\,, \sqrt{3}/2\,, 1\}$, while the slope of the self-interaction potential has been fixed for all the models to $\alpha = 0.08$. For each model the background and the linear perturbations evolutions have been computed beforehand by numerically integrating Eqs.~\ref{klein_gordon}-\ref{friedmann} and Eqs.~\ref{gf_plus}-\ref{gf_minus}, respectively, starting from symmetric (i.e. $\Omega _{+} = \Omega _{-}$) and adiabatic (i.e. $\Omega _{+}\delta _{+} = \Omega _{-}\delta _{-}$) initial conditions at very high redshift ($z\approx 10^{7}$) in order to get the proper expansion history $H(z)$ and growth factor $D(z)\equiv (1+z)(\Omega _{+}\delta_{+} + \Omega _{-}\delta _{-})/(\Omega _{+} + \Omega _{-})$. These have been used to rescale the amplitude of density perturbations in the simulations initial conditions, even if the deviation from the standard growth rate and expansion history can be considered negligible for the range of couplings under investigation. In fact, thanks to the very efficient background screening of McDE models \citep[][]{Piloyan_etal_2013} the expansion history of all the cosmologies presented in this work is consistent with a standard $\Lambda $CDM evolution with WMAP7 cosmological parameters \citep[][see Table~\ref{tab:parameters}]{wmap7} at the sub-percent level, with a maximum relative deviation from the $\Lambda $CDM Hubble function of $0.07\%$ at $z\sim 0.2$. Similarly, the linear growth factor of all the models is consistent with a $\Lambda $CDM growth rate having a maximum deviation of $\approx 0.05\%$.\\

Initial conditions have been generated using a version of the public {\small N-GenIC} code specifically modified
to account for multiple CDM particle species with generic abundances and relative perturbations amplitudes. 
The procedure goes as follows: first, a random-phase realisation of the total matter power spectrum -- which was assumed to be the matter power spectrum of a $\Lambda $CDM cosmology with WMAP7 cosmological parameters as computed with the Boltzmann code {\small CAMB} \citep[][]{CAMB} -- is generated for a set of $512^{3}$ particles by computing displacements from a {\em glass} \citep[][]{White_1994,Baugh_etal_1995} homogeneous distribution according to Zel'dovich approximation \citep[][]{Zeldovich_1970}, and its amplitude is adjusted to the starting redshift of the simulation using the total growth factor $D(z)$ computed as described above; then, before applying the displacements each particle is split into two particles with a mass ratio corresponding to the desired relative abundance of the two CDM species (for our symmetric models, this ratio is then simply $\Omega _{+}/\Omega _{-}=1$); finally, the displacement computed for each original particle is applied to its corresponding particle pair according to the desired relative amplitude of perturbations: the particles are moved in opposite {\em randomly chosen} directions with respect to their original position by a fraction of the total displacement corresponding to the share of the total perturbations amplitude of the two different CDM species. In other words, the original displacement $\psi _{i}$ computed for the $i$-th particle of the total CDM field is assigned to the corresponding particles of the two CDM species as $\psi _{i\,,\pm} = \Omega _{\pm }\delta _{\pm}\psi _{i}/\Omega _{\rm CDM}$. These particles are then moved by a $\psi _{i\,,\pm }$ displacement in opposite directions along a randomly chosen line. This procedure ensures that the desired relative perturbations amplitude (in our adiabatic case simply $\Omega _{+}\delta _{+}/\Omega _{-}\delta _{-}=1$) is correctly represented in the initial conditions, and that no preferred direction is introduced in the system.
The random seed adopted for the initial realisation of the desired power spectrum has been kept the same for all the simulations so to ensure the possibility of directly comparing structures on an individual basis.
\\

For each simulation snapshot we have identified CDM halos through a Friends-of-Friends (FoF) algorithm with a linking length of $0.2$ times the mean inter-particle separation, by linking together particles of both CDM species without any distinction of particle type. This procedure provides a catalog of CDM structures identified only through the CDM particle density, consistently with what could be directly tested as no distinction between the two different CDM particle types is observationally accessible. We have stored  groups with at least $32$ particles, corresponding to a minimum mass of $\approx 10^{10}$ M$_{\odot }/h$. 

Additionally, we have also employed the {\small SUBFIND} algorithm \citep[][]{Springel_etal_2001} embedded in {\small GADGET-3} to identify gravitationally bound substructures within FoF groups, based on the standard gravitational potential of the CDM field. Clearly, such procedure is strictly rigorous only for the uncoupled case $|\beta |=0$, as for coupled models the binding energy of particles is modified by the fifth-force corrections of Eq.~\ref{gf_plus}-\ref{gf_minus}. Nonetheless, for small coupling values the inaccuracy in the determination of the position of the potential minimum of bound halos (which is the only information from {\small SUBFIND} that we use in the present work) is expected to be negligible.

For all our simulations snapshots we have then also computed the total CDM density field on a $2048^{3}$ cartesian grid through a {\em Cloud-in-Cell} (CIC) mass assignment algorithm, and all the images of the CDM distribution presented in this work (as e.g. Figs.~\ref{fig:lss}, \ref{fig:zoom}, \ref{fig:closeup}) have been produced by mapping the logarithm of such density field to the same brightness scale calibrated on the uncoupled simulation output at z=0.

\section{Results}
\label{sec:results}

We now move to describe the main results of our numerical investigation of McDE cosmologies, starting from the visual inspection of the CDM distribution
at small scales, and then moving to the more quantitative determination of statistical and structural properties of CDM structures. To avoid confusion in the plots
we will not display all the 7 simulations of our suite in most of the figures, focusing on a representative subsample of 5 runs ($|\beta |=\{ 0\,, 1/2\,, 3/4\,, \sqrt{3}/2\,, 1\}$)
as the remaining models ($|\beta |=\{7/10\,, 8/10\}$) are found to add no significant information to the final results and have been mainly used to test the transition between the $|\beta |<\sqrt{3}/2$ and the $|\beta |>\sqrt{3}/2$ coupling regimes.

\subsection{Dark matter distribution}
\begin{figure*}
\begin{center}
\includegraphics[width=3.5in]{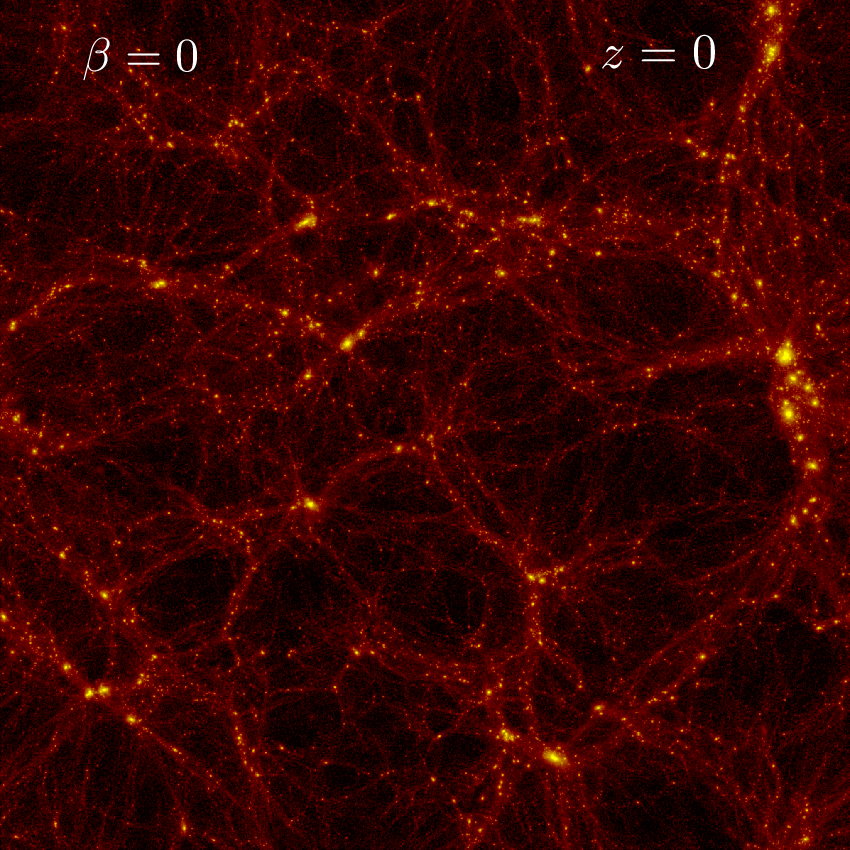}\\
\includegraphics[width=3.5in]{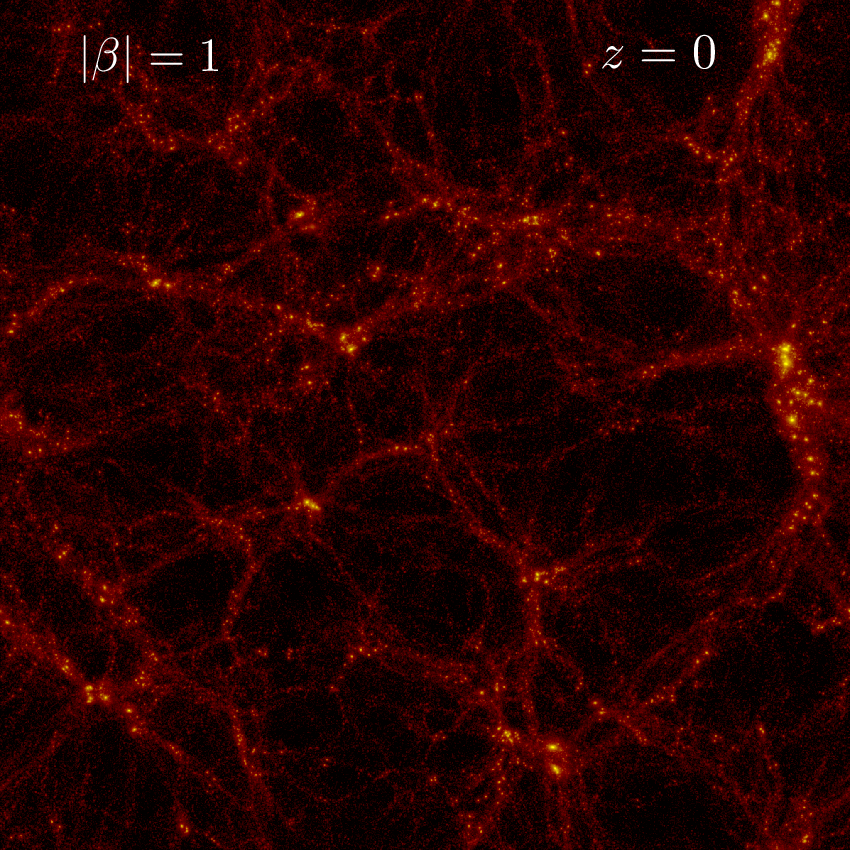}
\end{center}
\caption{Large-scale CDM distribution over the whole simulation box in the two extreme cosmologies considered in the present work, $|\beta |=0$ ({\em top}) and $|\beta |=1$ ({\em bottom}). {\em (High-resolution figures available online)}}
\label{fig:lss}
\end{figure*}

We start our discussion by comparing the spatial distribution of the total CDM fluid in the different models. In Fig.~\ref{fig:lss} we display 
the CDM density field in a slice of thickness 30 Mpc$/h$ through the whole simulation box in the two most extreme models under investigation, i.e. the uncoupled case $|\beta |=0$ ({\em top}) and the unitary coupling $|\beta |=1$ ({\em bottom}). As one can see from the figure, no major difference is present in the large-scale topology of the CDM distribution between the two models, with overdense regions and voids appearing in the same locations and with a comparable geometry. However, the CDM density field looks more ``smeared" 
and diffuse in the coupled case as compared to the uncoupled cosmology, with a clearly lower density contrast in the peaks of the overdense regions, and a less sharp 
transition between filaments and voids. We note that all such features might be in principle detectable in terms of weak and strong gravitational lensing observables which might provide a way to rule out unitary couplings in the context of McDE scenarios. 

Furthermore, by looking at the most overdense regions in the two figures, one can easily notice
how the same structures appear to be populated with a larger number of substructures in the coupled case as compared to the uncoupled model. This is the first direct evidence
of the halo fragmentation process that takes place at low redshifts in McDE cosmologies with coupling values $|\beta |\geq \sqrt{3}/2$ \citep[as already hinted by][]{Baldi_2013} and that will be discussed in detail below. In general, the direct comparison of Fig.~\ref{fig:lss} shows how at large scales even the most extreme McDE model of our sample
has mainly the effect of smoothing the total density field and of increasing the number of halo substructures, without compromising too dramatically the large-scale shape of the cosmic web.\\
\begin{figure*}
\includegraphics[width=2.7in]{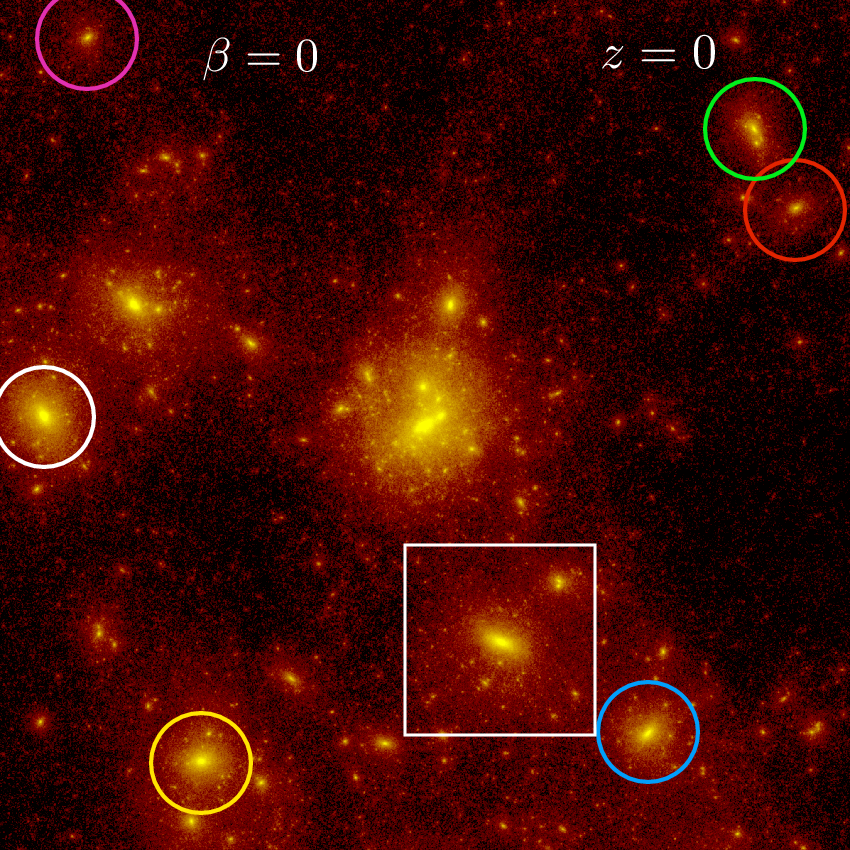}
\includegraphics[width=2.7in]{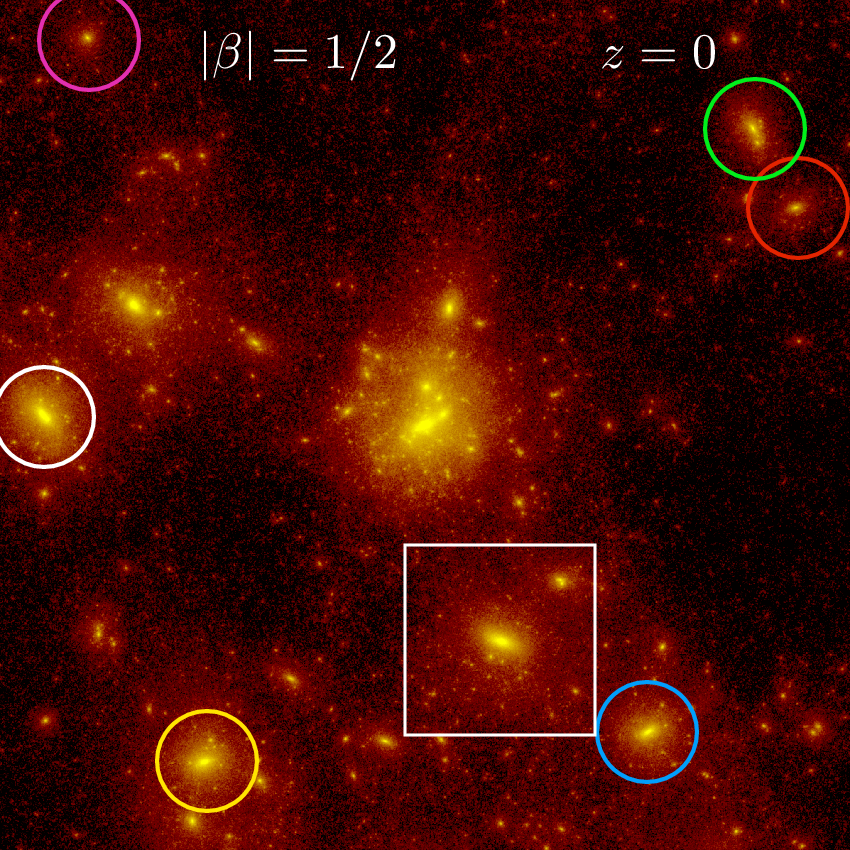}
\includegraphics[width=2.7in]{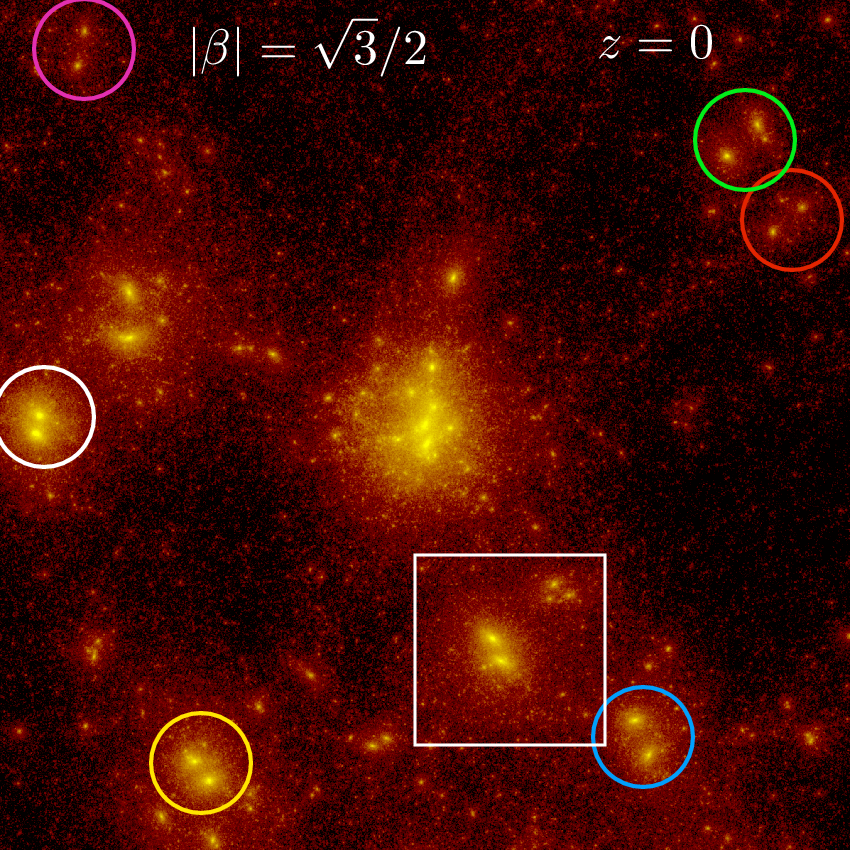}
\includegraphics[width=2.7in]{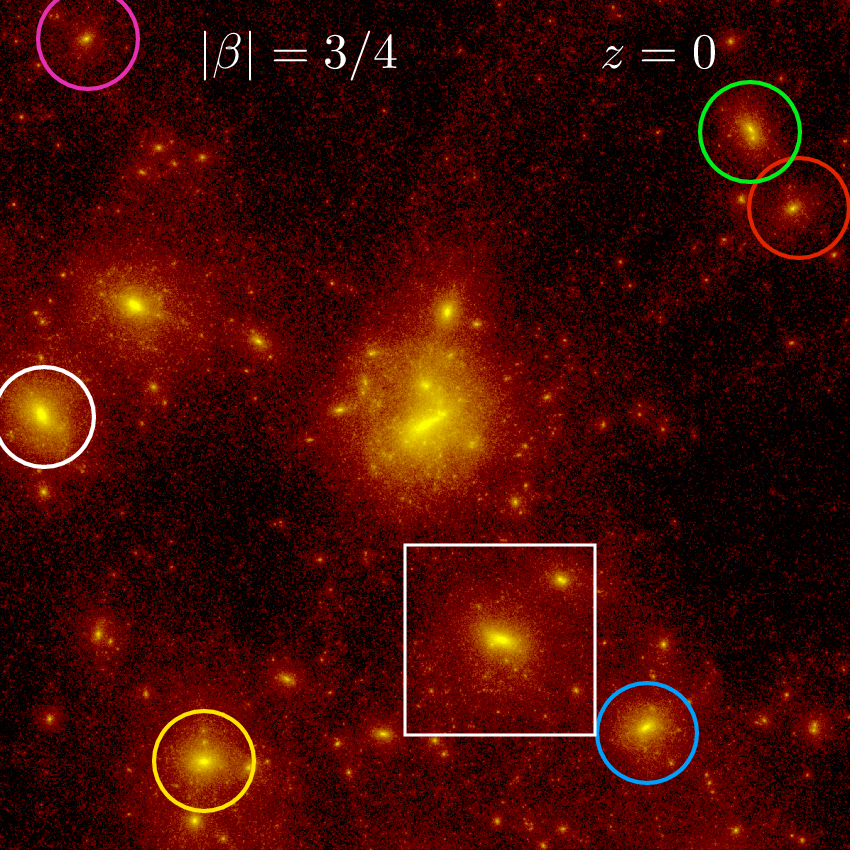}
\caption{Zoom on the most massive halo of the simulation in four different McDE models with clockwise increasing coupling starting from the uncoupled case in the upper left plot. The size of each image is $10\times 10$ Mpc$/h$, and corresponding objects have been highlighted with coloured circles and a large square. {\em(High-resolution figures available online)}}
\label{fig:zoom}
\end{figure*}

\begin{figure*}
\includegraphics[width=2.7in]{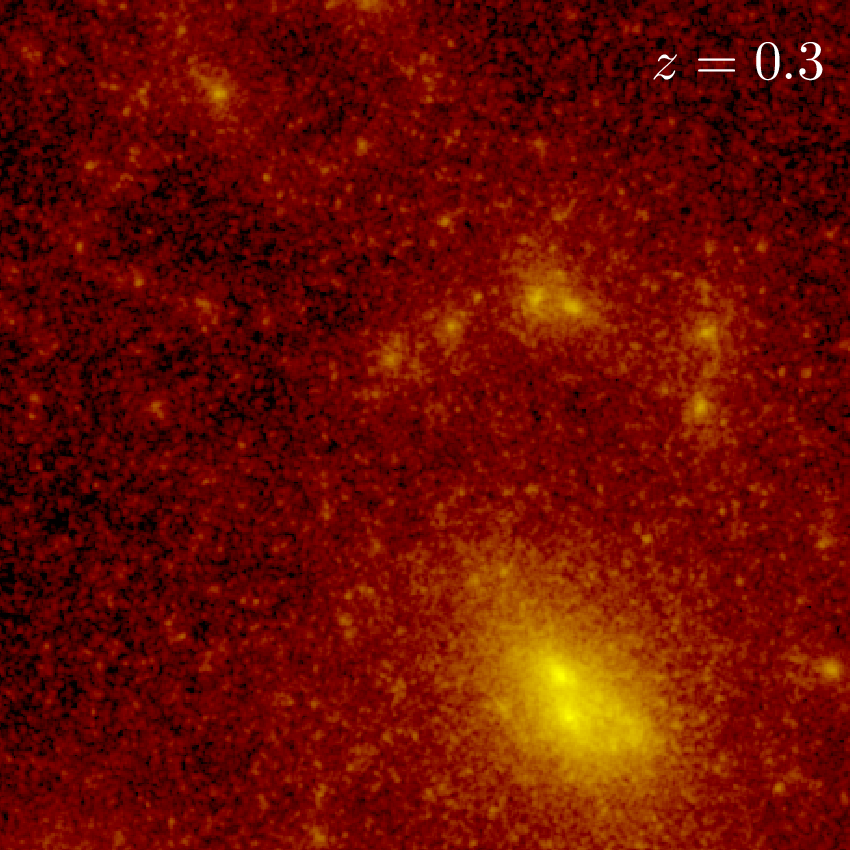}
\includegraphics[width=2.7in]{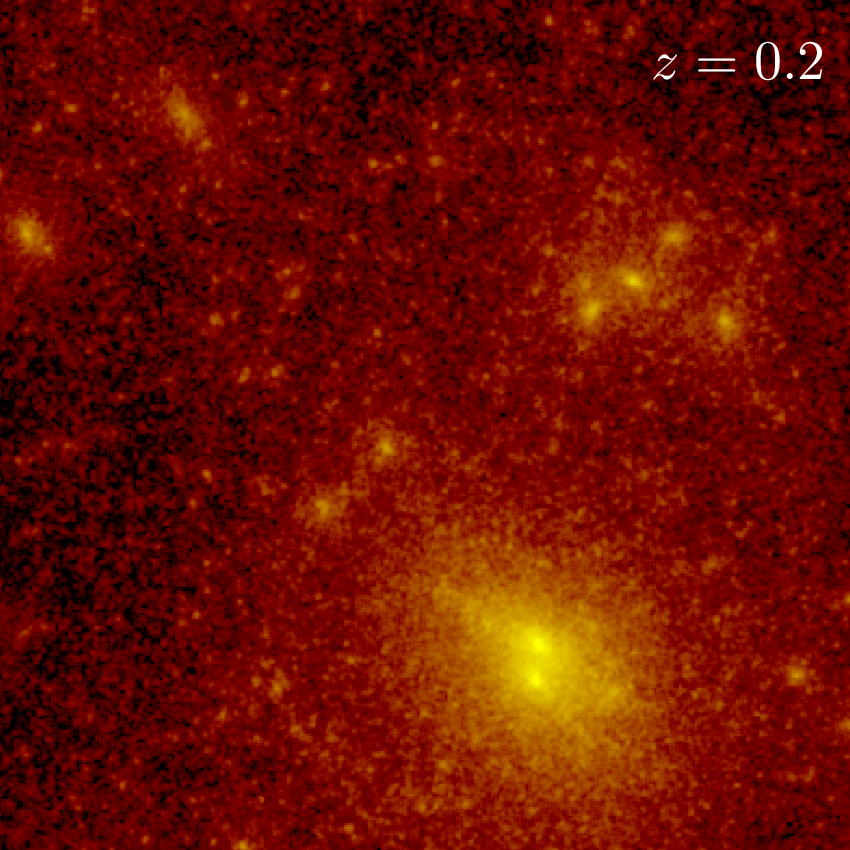}\\
\includegraphics[width=2.7in]{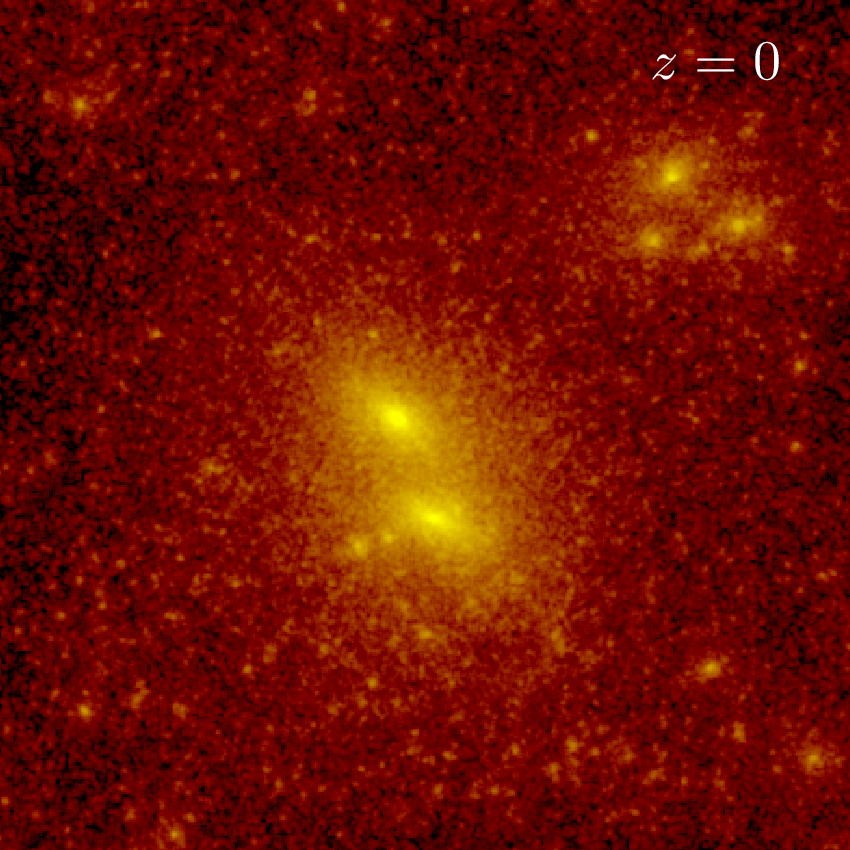}
\includegraphics[width=2.7in]{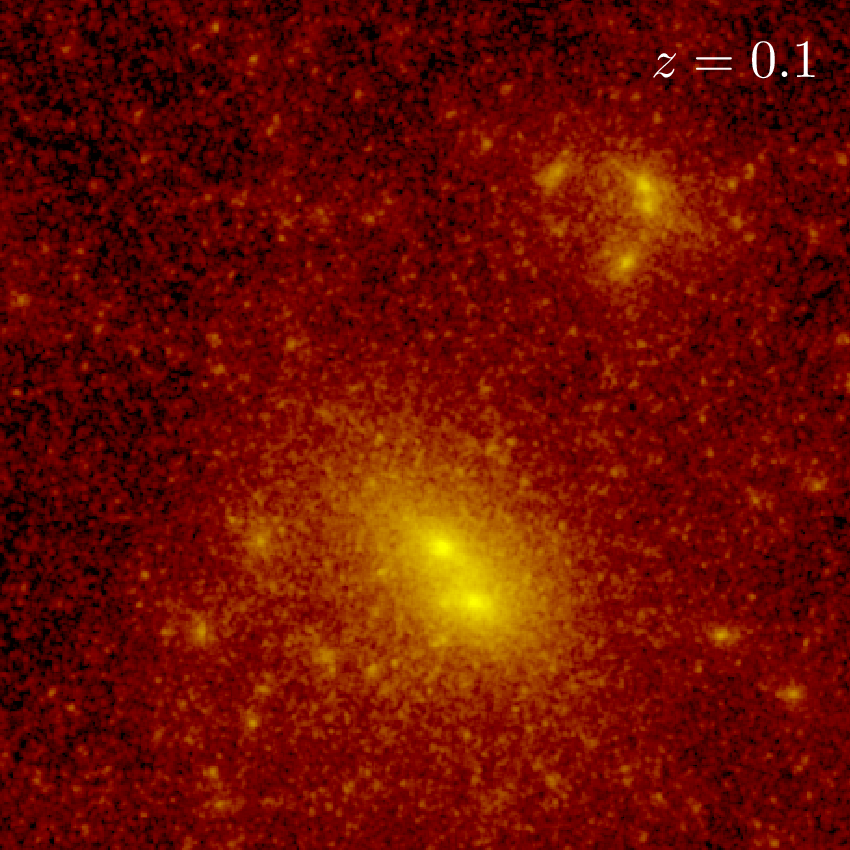}
\caption{The evolution of the infalling halo highlighted by the white square in Fig.~\ref{fig:zoom}  for the $|\beta | = \sqrt{3}/2$ model at different decreasing redshifts, displayed clockwise starting from the upper left panel. The fragmentation of the halo in two smaller halos can be clearly seen evolving in time. Also, the trajectories of the two fragmented objects progressively deviate from each other, thereby providing a direct evidence of the violation of the weak equivalence principle in McDE cosmologies. {\em (High-resolution figures available online)}}
\label{fig:closeup}
\end{figure*}

Moving to smaller scales, we can then start to notice a series of more peculiar effects of McDE cosmologies that could not be explored with previous low-resolution simulations.
In Fig.~\ref{fig:zoom} we show the density field of the total CDM fluid in a slice of $10\times 10$ Mpc$/h$ centered on the most massive halo identified in the simulations at $z=0$ for 4 different McDE models with increasing coupling, displayed clockwise starting from the uncoupled case in the upper-left panel. Besides the central massive structure, several other objects orbiting around the main halo or moving toward it along filaments appear in the figure, and 7 of them have been highlighted by 6 small coloured circles and one larger white square. The latter defines the field of view that is displayed in Fig.~\ref{fig:closeup}, and that will be discussed below.
 
As one can see by comparing individual structures, the main effect of the DE-CDM multiple interaction up to $|\beta |=3/4$
appears to be a smoothing of the density field around collapsed objects resulting in a more diffuse CDM halo surrounding the main density peaks. This is particularly evident by comparing the density distribution around the central massive object and around some other smaller structures as e.g. those highlighted by the white or the blue circles located in the center-left and bottom-right parts of the figures, respectively. The most interesting effect, however, shows up for the gravitational coupling case $|\beta |=\sqrt{3}/2$ displayed in the bottom-left panel of Fig.~\ref{fig:zoom}, for which it is easy to notice that all the main structures highlighted in the figure have split into two smaller objects separated roughly by the same projected distance. For such model, also the central halo appears to have been fragmented into a bunch of smaller objects, highlighting the presence of several substructures within the original main halo that were not visible in the other images due to resolution limits of the density maps. 

This behaviour is particularly interesting for at least two reasons: first, it directly shows the onset of the halo fragmentation process expected to occur in McDE models \citep[][]{Baldi_2013}, thereby confirming how the coupling threshold for such process is given by the gravitational coupling barrier $|\beta |=\sqrt{3}/2$ (as also the $|\beta |=8/10$ run -- which is not displayed in Fig.~\ref{fig:zoom} but has been included in the supplementary material -- did not show any evidence of halo splitting); second, the visual inspection of Fig.~\ref{fig:zoom} demonstrates that the 
halo fragmentation process is not necessarily triggered by the repulsive fifth-force between CDM particles of opposite species, but can also be a consequence (at least for the gravitational coupling case) of the
different friction terms of Eqs.~\ref{gf_plus}-\ref{gf_minus}. In fact, it is important to remind here they the $|\beta |=\sqrt{3}/2$ model
displayed in Fig.~\ref{fig:zoom} corresponds to the unique point in the McDE parameter space for which the repulsive fifth-force corrections exactly balance the standard gravitational attraction, such that for this coupling value CDM particles of opposite type do not interact with each other in any way. Therefore, the split of individual bound objects into pairs of CDM halos cannot be the consequence of the repulsion between CDM particles of opposite type but must instead be associated to the differential friction acting on the two CDM particle species. Such conclusion is also supported by the observation that most of the fragmented halo pairs seem to be aligned along the direction of filaments pointing to the central massive structure, that represent the most likely directions of motion of the original parent halos along which the differential friction is expected to act.

\begin{figure}
\begin{center}
\includegraphics[width=2.5in]{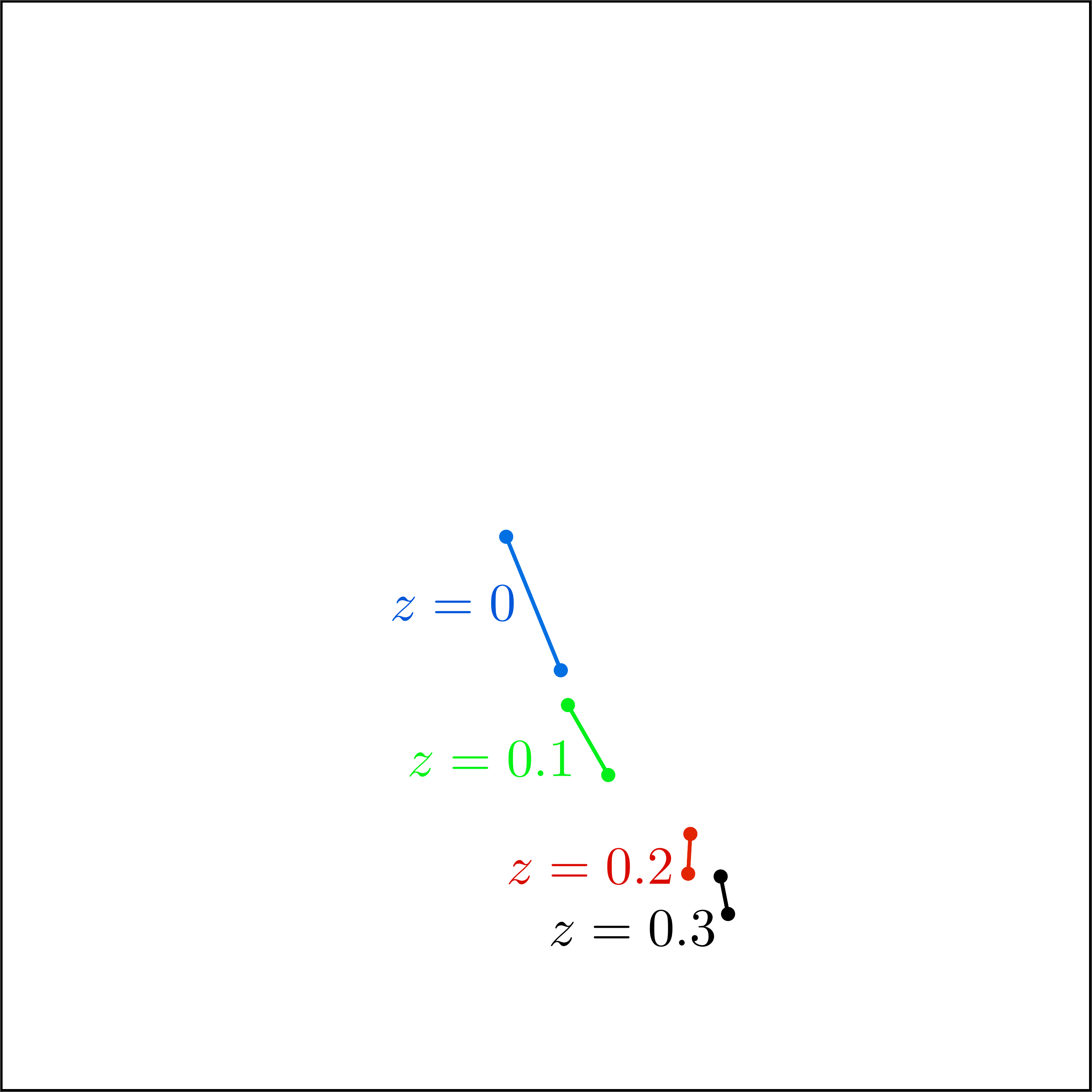}
\end{center}
\caption{The projected trajectory of the halo pair highlighted in the square field of view of Fig.~\ref{fig:closeup}. The halo separation grows along the trajectory of the original parent halo.}
\label{fig:closeup_trajectory}
\end{figure}

To investigate in further detail such effect, and to provide support to the latter conclusion, in Fig.~\ref{fig:closeup}, we show a zoom of the field of view highlighted by the white square in the $|\beta |=\sqrt{3}/2$ plot of Fig.~\ref{fig:zoom} at four different redshifts $z=\{ 0.3\,, 0.2\,, 0.1\,, 0\}$ in order to 
highlight how the separation between the two halos arising from the fragmentation of the same original structure evolves in time. Indeed, the figure clearly shows how the separation of the two objects grows in time, thereby providing a direct evidence of the violation of the weak equivalence principle in McDE models: two test objects moving in the same gravitational field and with the same initial infall velocity are found to have different dynamical trajectories. Also, the comparison of Fig.~\ref{fig:closeup}
qualitatively confirms that the two objects  are displaced  along their trajectory of motion, and that their separation along such trajectory grows in time, consistently with the effect of a different friction term. The same result, confirming our interpretation of the onset of halo fragmentation for the gravitational coupling $|\beta |=\sqrt{3}/2$, is also more quantitatively expressed by Fig.~\ref{fig:closeup_trajectory}, displaying within the same field of view the projected separation of the halo pair and its projected trajectory at different redshifts. As one can see from the figure, as time goes by the two fragmented halos move in the same direction but their separation along the projected trajectory grows in time.

\subsection{Halo fragmentation}

\begin{figure*}
\begin{center}
\caption{The virial abundance ratio of Eq.~(\ref{abundance_ratio}) of the 1000 most massive  CDM halos at different redshifts ({\em left}), and its statistical distribution ({\em right}).}
\includegraphics[scale=0.25]{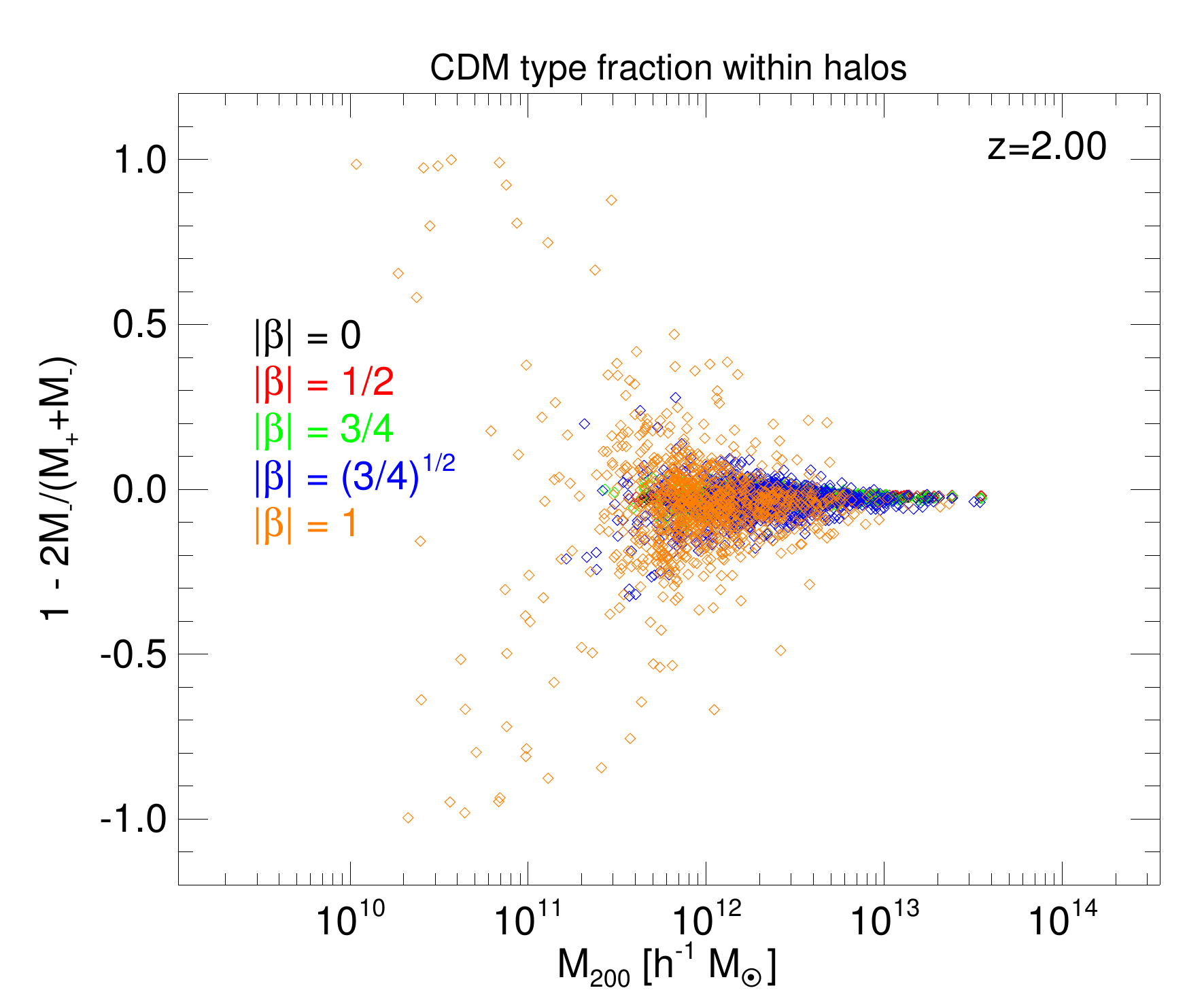}
\includegraphics[scale=0.25]{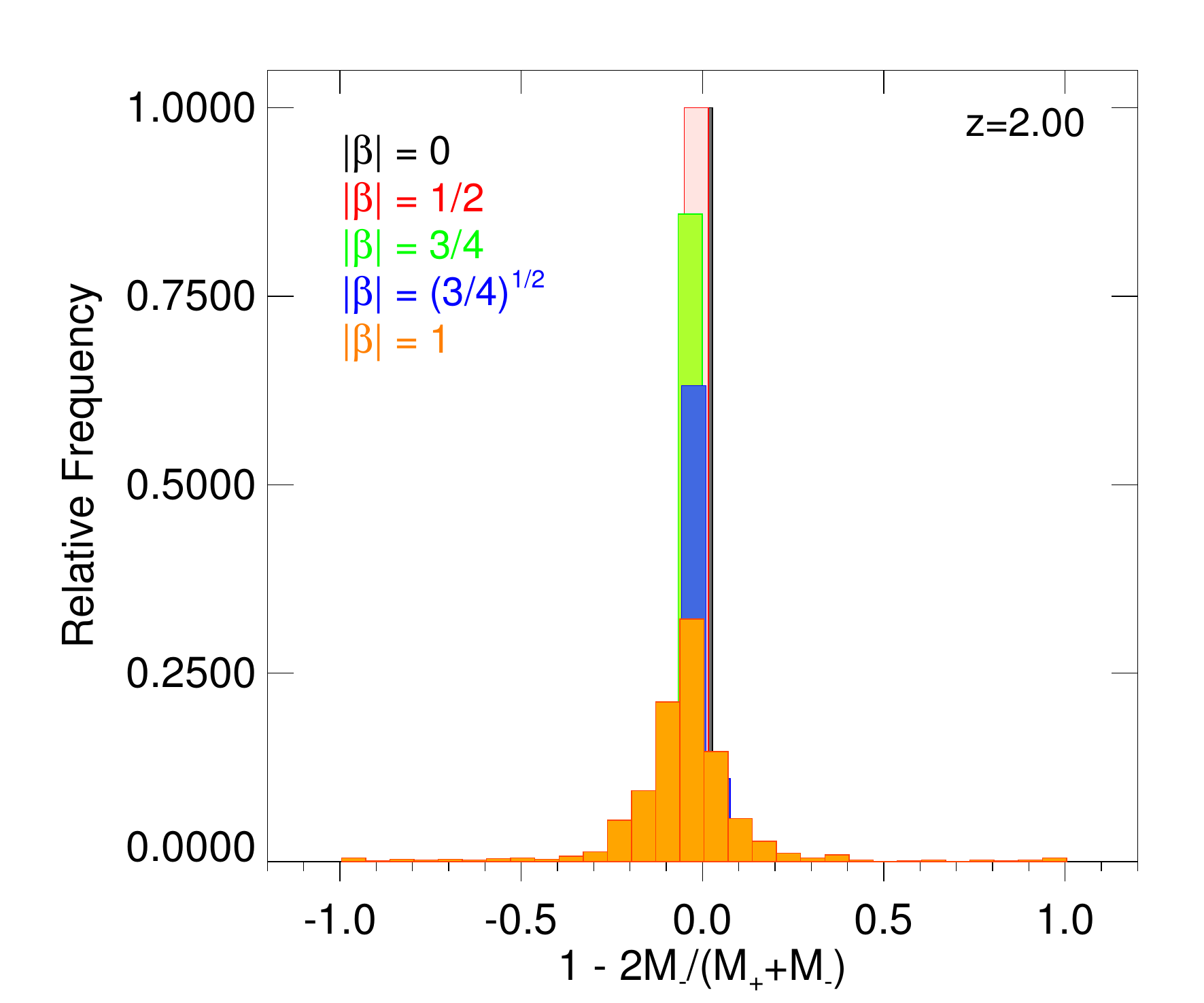}\\
\includegraphics[scale=0.25]{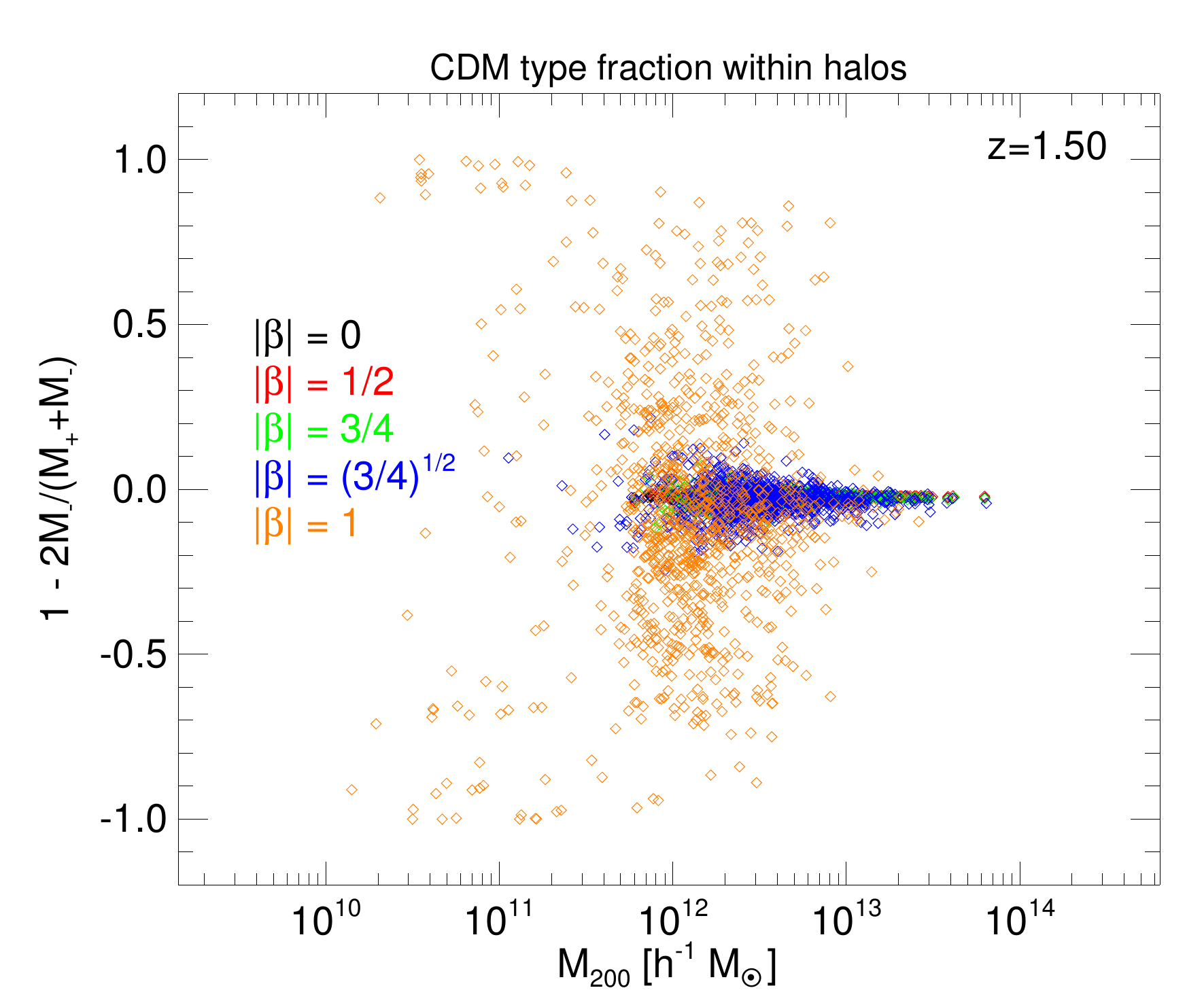}
\includegraphics[scale=0.25]{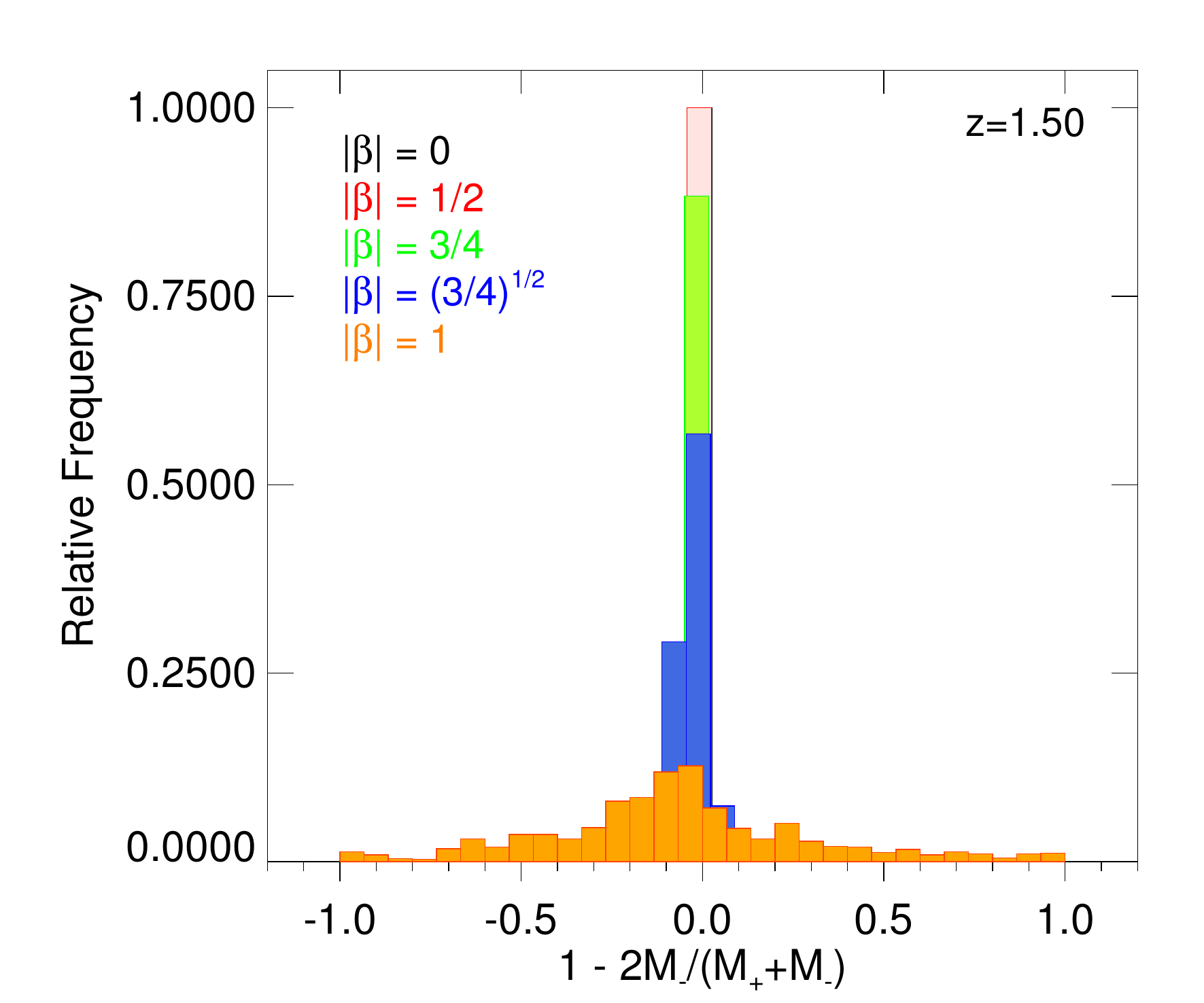}\\
\includegraphics[scale=0.25]{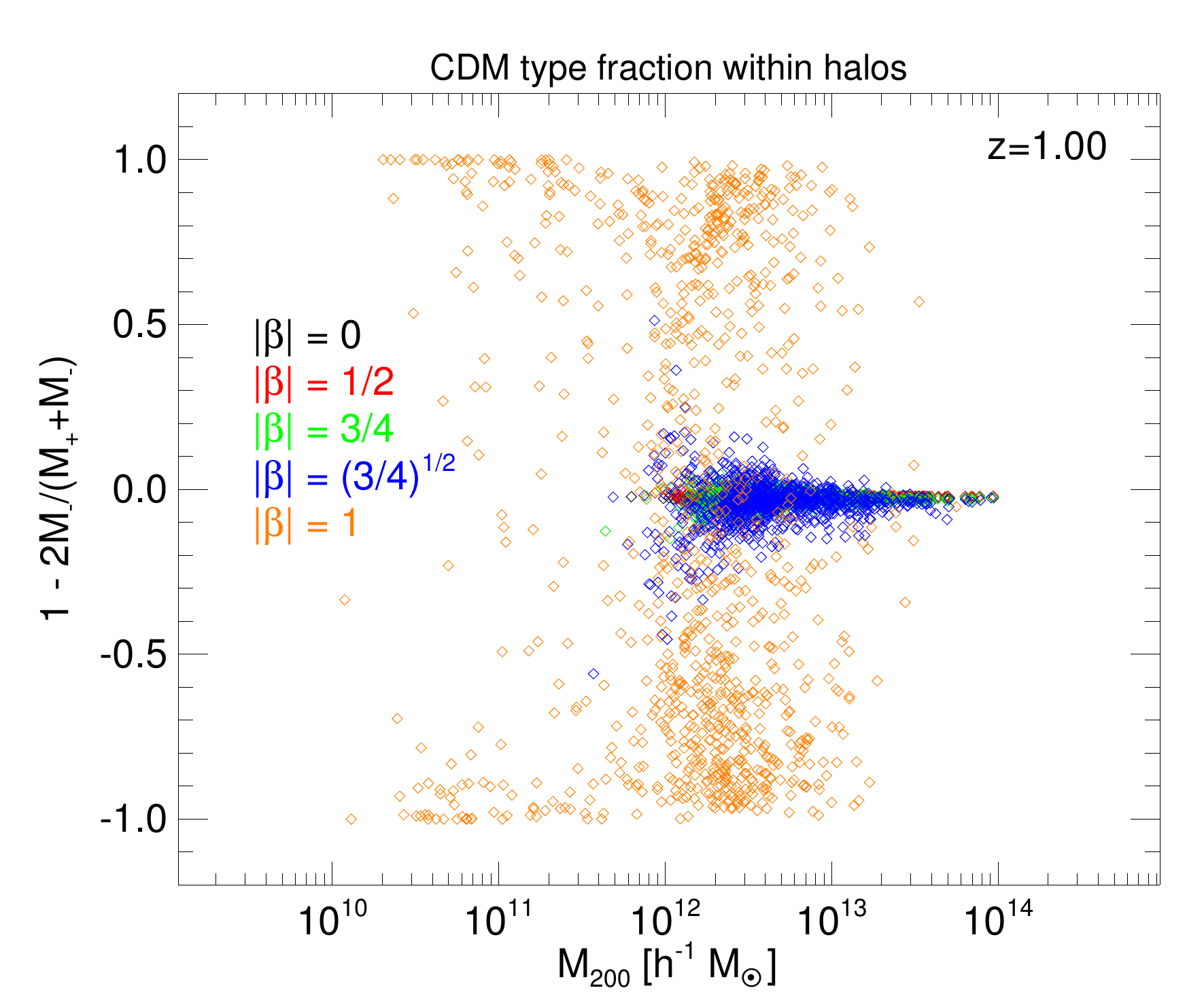}
\includegraphics[scale=0.25]{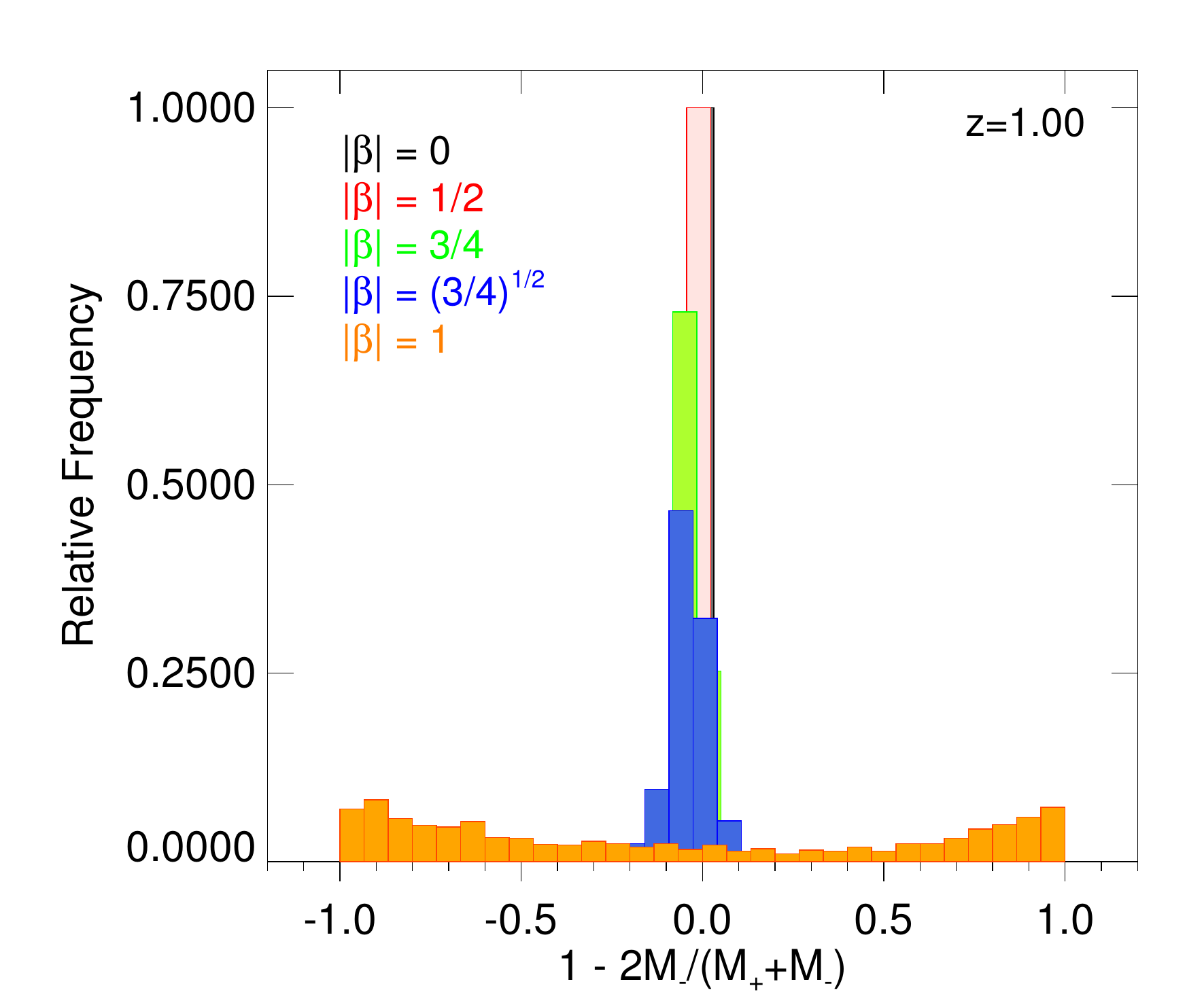}\\
\includegraphics[scale=0.25]{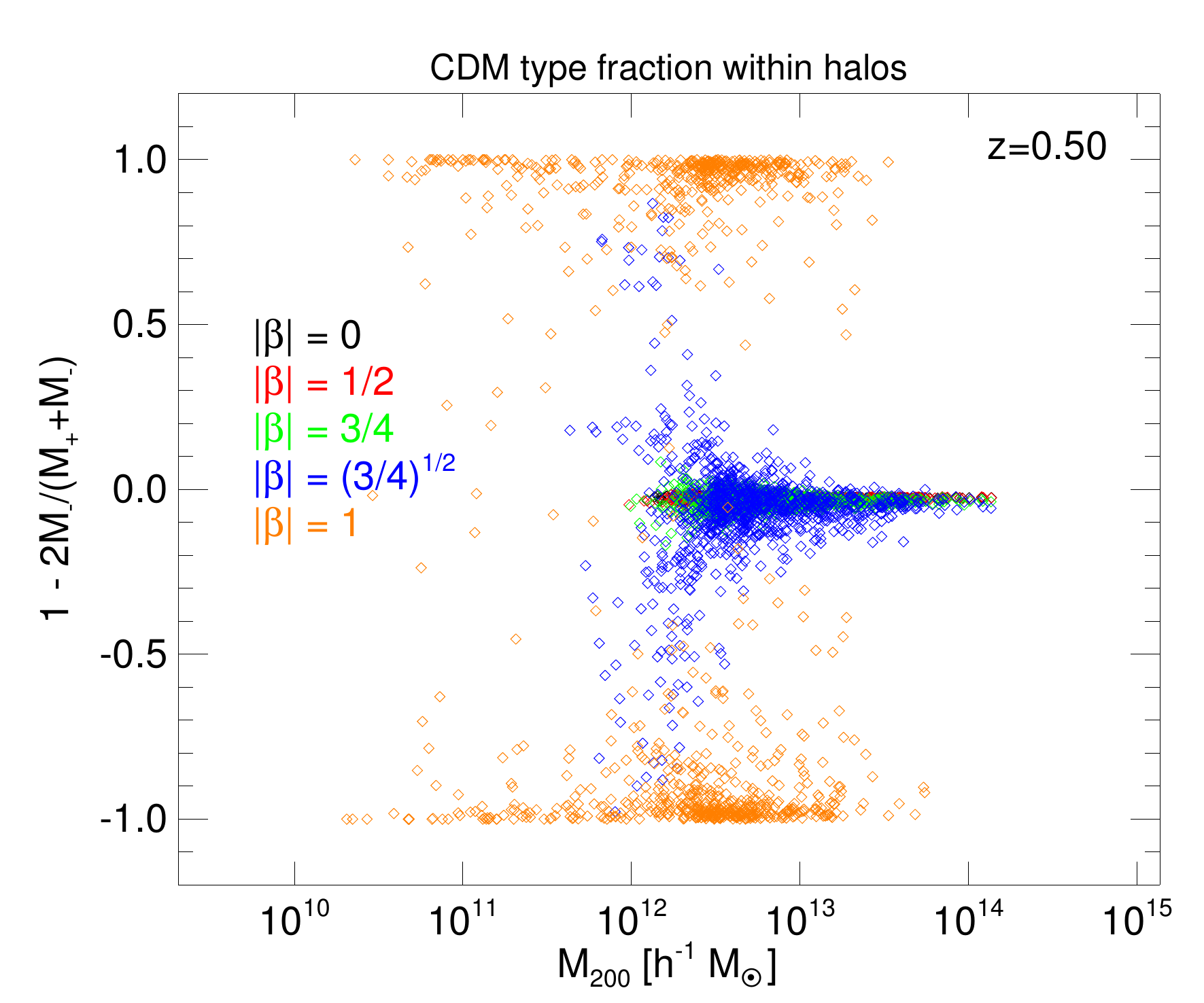}
\includegraphics[scale=0.25]{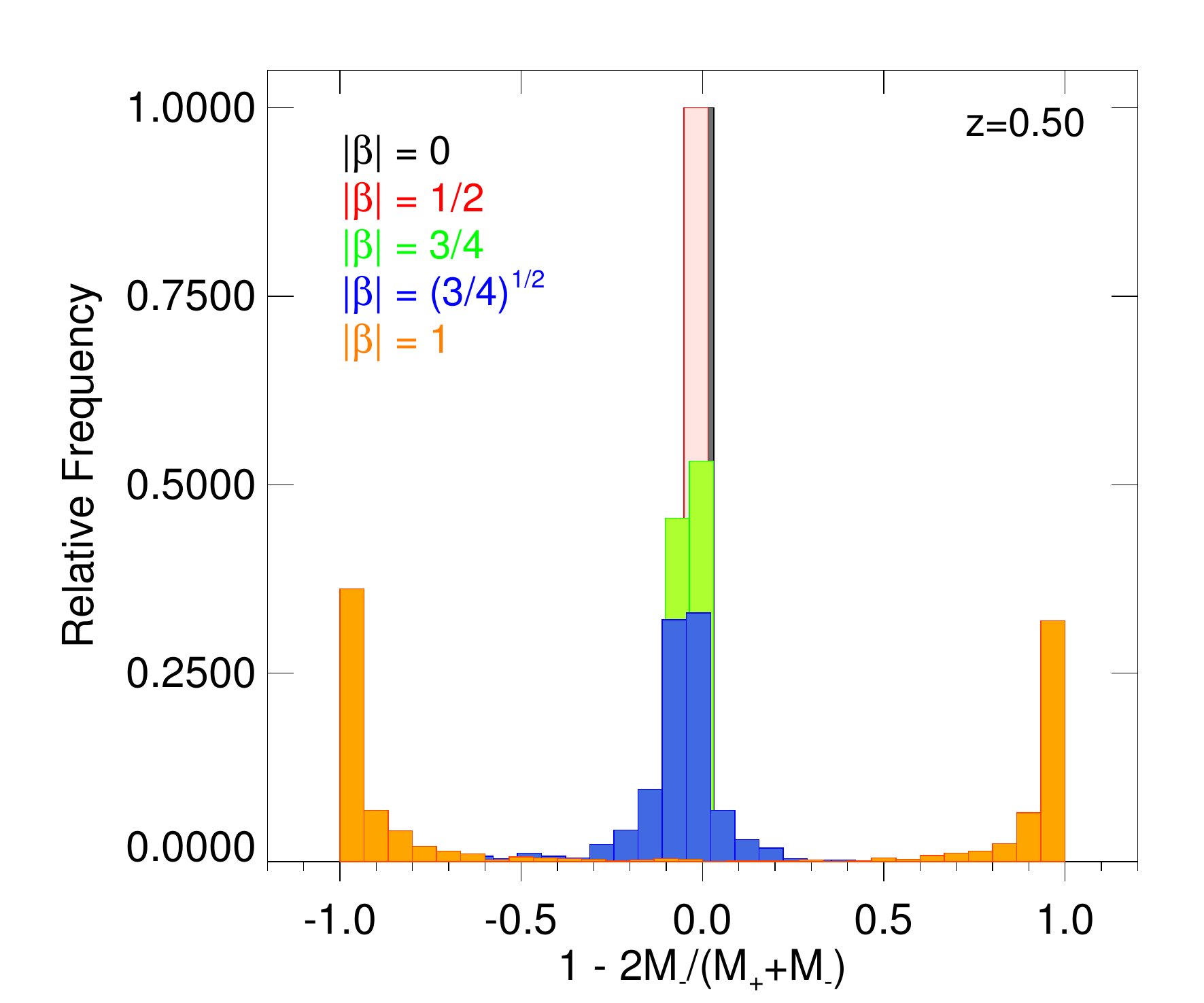}\\
\includegraphics[scale=0.25]{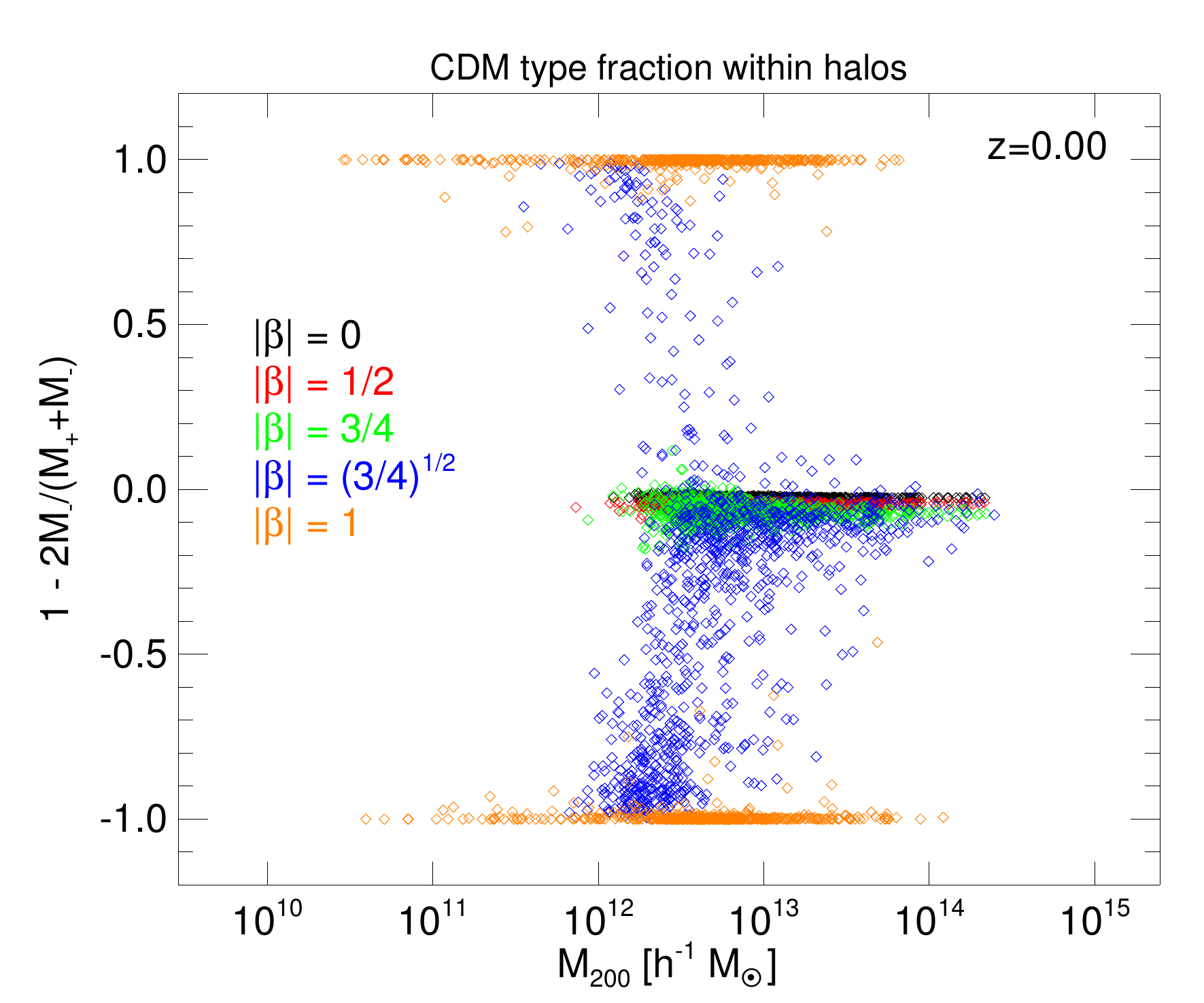}
\includegraphics[scale=0.25]{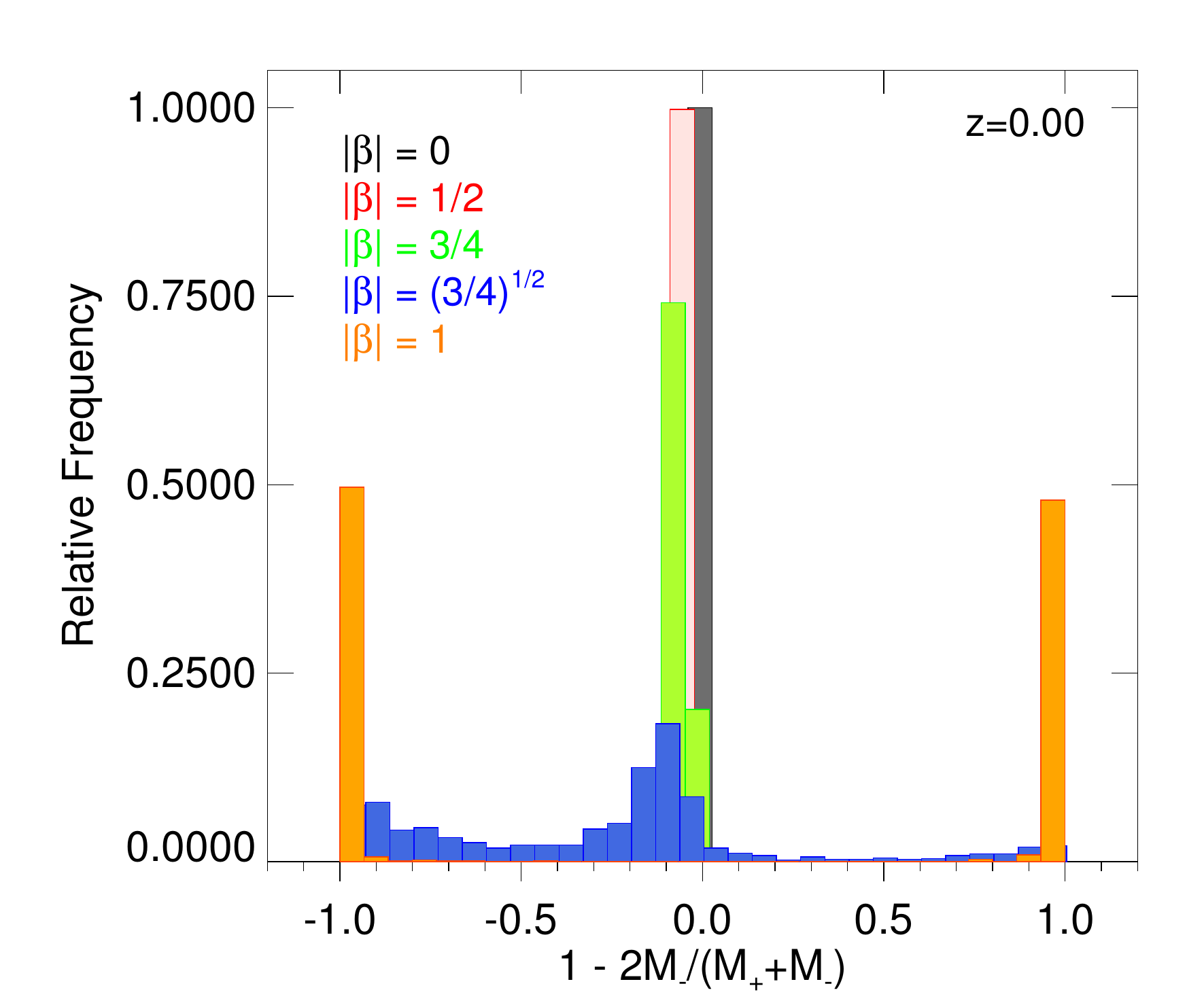}
\label{fig:type_fraction}
\end{center}
\end{figure*}

In the present Section we investigate in some further detail how the halo fragmentation process responsible of breaking the screening mechanism for 
large coupling values evolves as a function of time and of the halo mass. 
We consider the 1000 most massive halos identified through the FoF algorithm in all our simulations, and
for each CDM halo in our sample
we compute the virial abundance ratio of the two different CDM species defined as:
\begin{equation}
1-\frac{2M_{-}}{M_{+}+M_{-}}\,,
\label{abundance_ratio}
\end{equation}
where the masses $M_{\pm }$ are defined as the mass in the two particle types contained within the virial radius $R_{\rm 200}$ of each halo. Such quantity
will then take the value of $-1$ for objects made only of negatively-coupled particles, and the value of $+1$ for objects made solely of positively-coupled particles.
Any other value in between indicates a mixture of particle types, with $0$ being the value corresponding to a perfect balance between the two species that is expected to hold for an uncoupled system.
In Fig.~\ref{fig:type_fraction} we display in the left plots the virial abundance ratio of Eq.~(\ref{abundance_ratio}) for all the halos in our sample and for the different cosmologies as a function of the halo mass, at different redshifts between $z=2$ and $z=0$. The various models are plotted in order of increasing coupling, such that the small coupling data points might be covered by the larger coupling ones. On the right plots, instead, we display for the same
redshifts the statistical distribution of the virial abundance ratio in the different models binned in 20 equally spaced bins over the allowed domain $[-1\,, 1]$.

As one can see from the plots, the halo fragmentation process indeed occurs only for coupling values equal or larger than the gravitational strength $|\beta |=\sqrt{3}/2$, for which we find
objects fully dominated by one single CDM particle type by $z=0$. The process occurs earlier and evolves faster for the unitary coupling model, that already
at $z=2$ has a distribution of the virial abundance ratio that spans the whole allowed range $[-1\,, 1]$, and that develops into a clearly bimodal distribution already before $z=0.5$. For the gravitational coupling model, instead, the halo fragmentation starts later and does never evolve to a fully bimodal distribution
before the present time, thereby showing a mixture of fragmented single-species halos and of multi-species halos, with the latter dominating the high-mass end of the distribution, and the former showing up at small masses. Also, the distribution of fragmented halos does not appear to be symmetric, with many more halos
being dominated by negatively-coupled particles. Such tendency will be also confirmed by the results shown in the middle panels of Fig.~\ref{fig:structural_properties} below. This asymmetry in the distribution of fragmented halos might indicate that for many of such structures the fragmentation process is not yet complete, and that the corresponding objects dominated by positively-coupled particles have not yet recollapsed to form a separate halo. 
In fact, such asymmetry does not appear anymore for the $|\beta | = 1$ case at low redshifts, when the halo fragmentation process has already completed, for which we find an equal abundance of objects dominated by the two CDM particle species.

Finally, for the McDE models with sub-gravitational couplings, the halo fragmentation does not occur at all, with the scalar interactions having only the effect of slightly broadening the distribution of the virial abundance ratio around the uncoupled value of $0$ that characterises all the models at high redshifts.\\

\subsection{The halo mass function}

\begin{figure*}
\begin{center}
\includegraphics[scale=0.37]{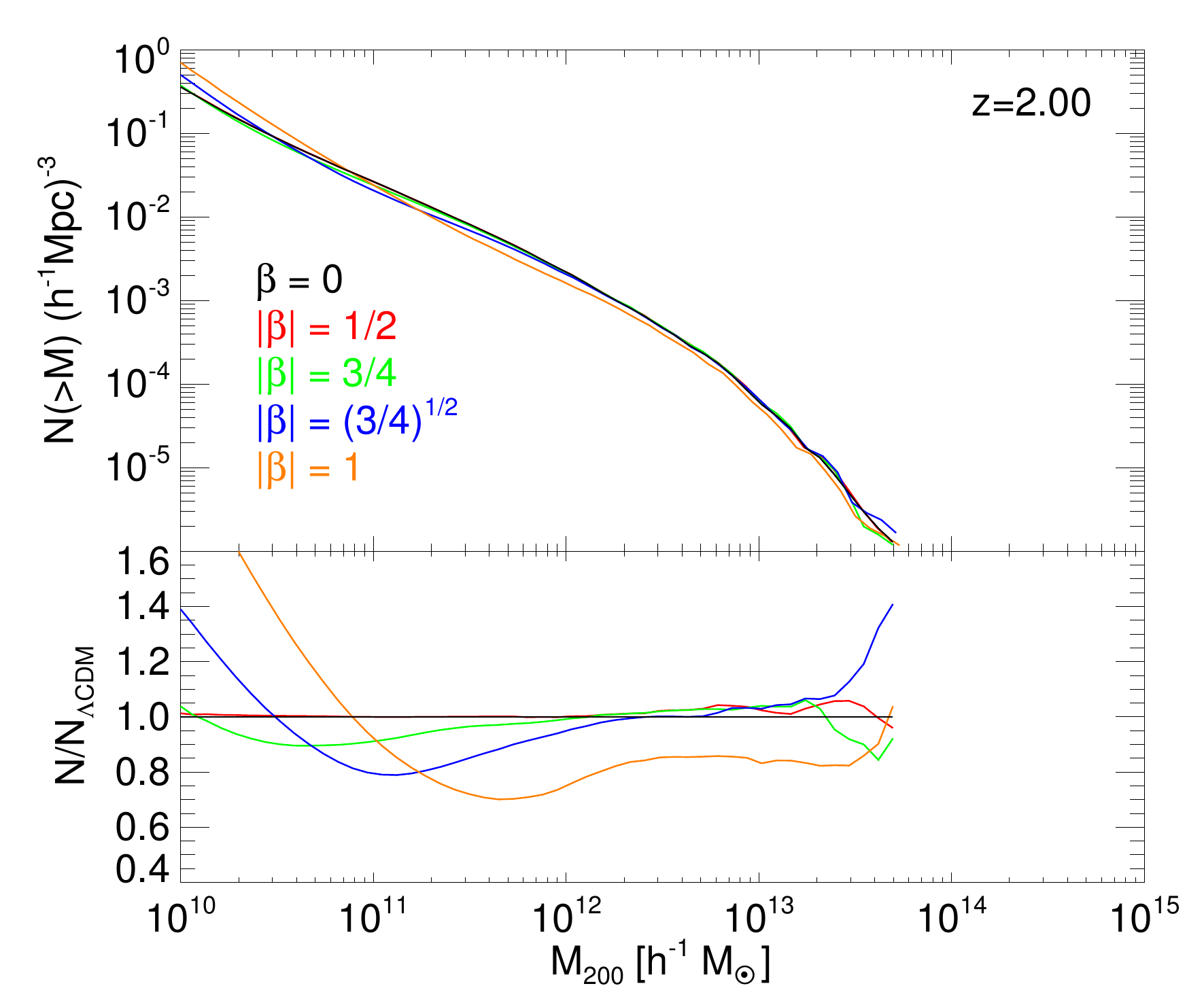}
\includegraphics[scale=0.37]{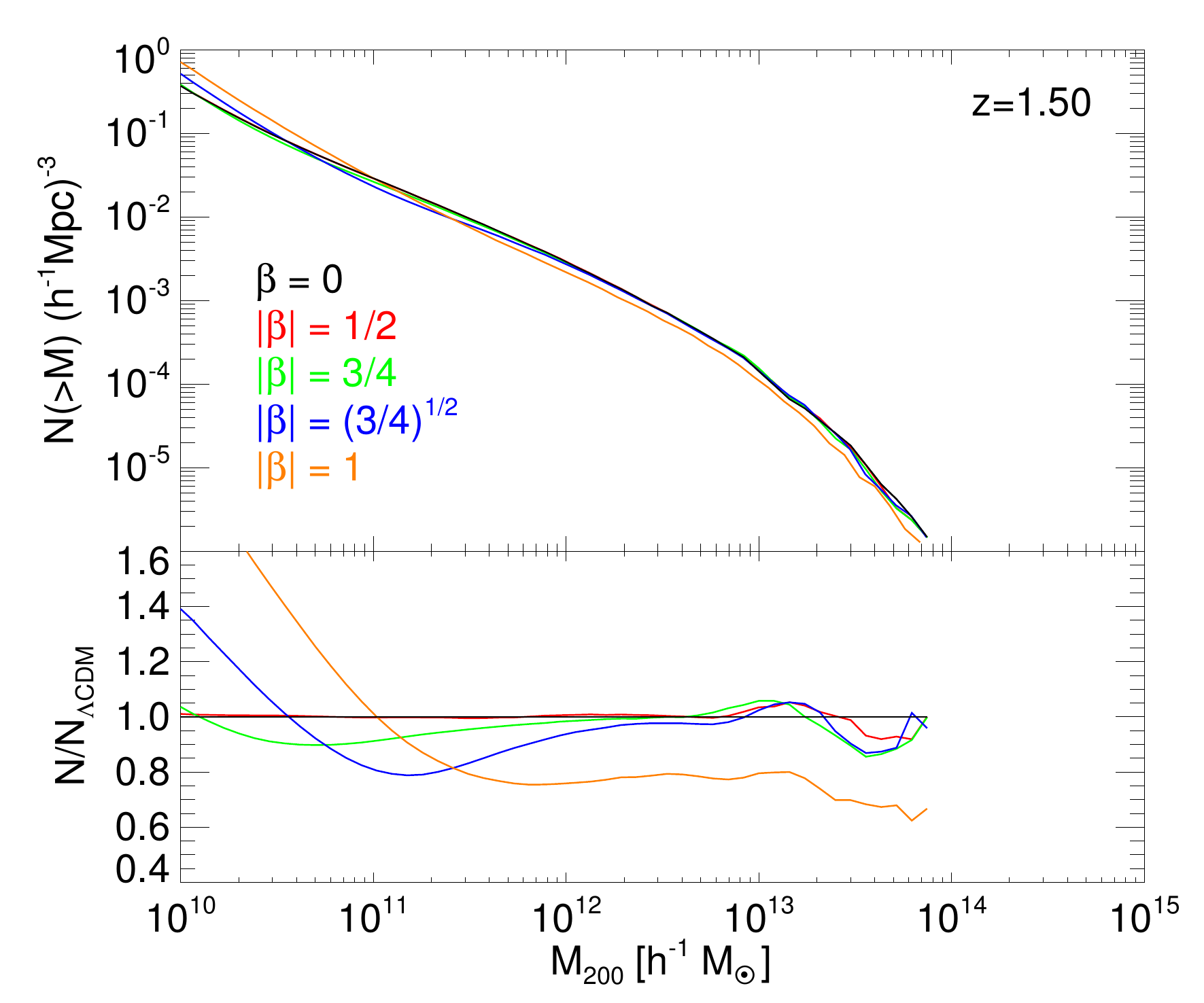}\\
\includegraphics[scale=0.37]{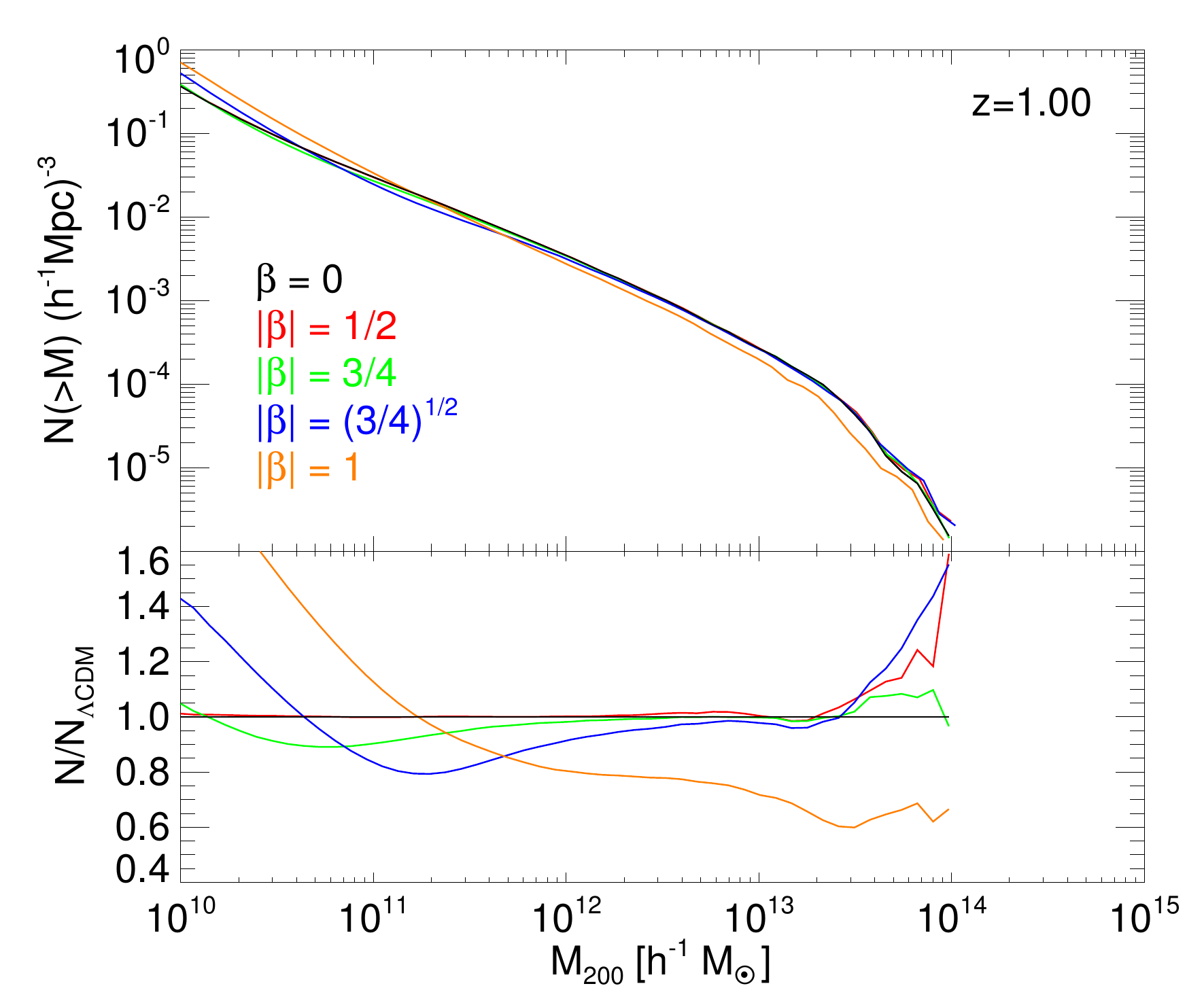}
\includegraphics[scale=0.37]{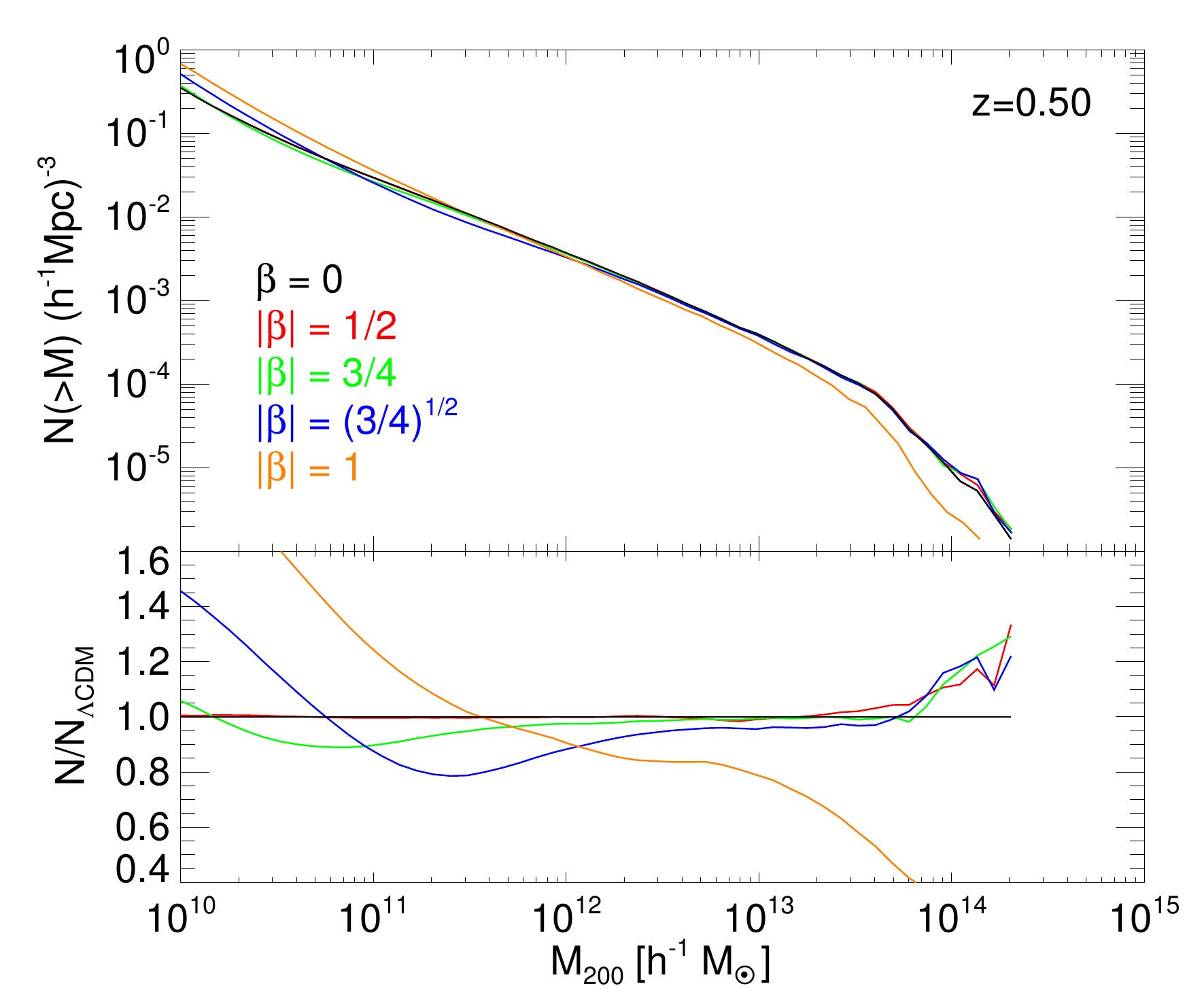}\\
\includegraphics[scale=0.37]{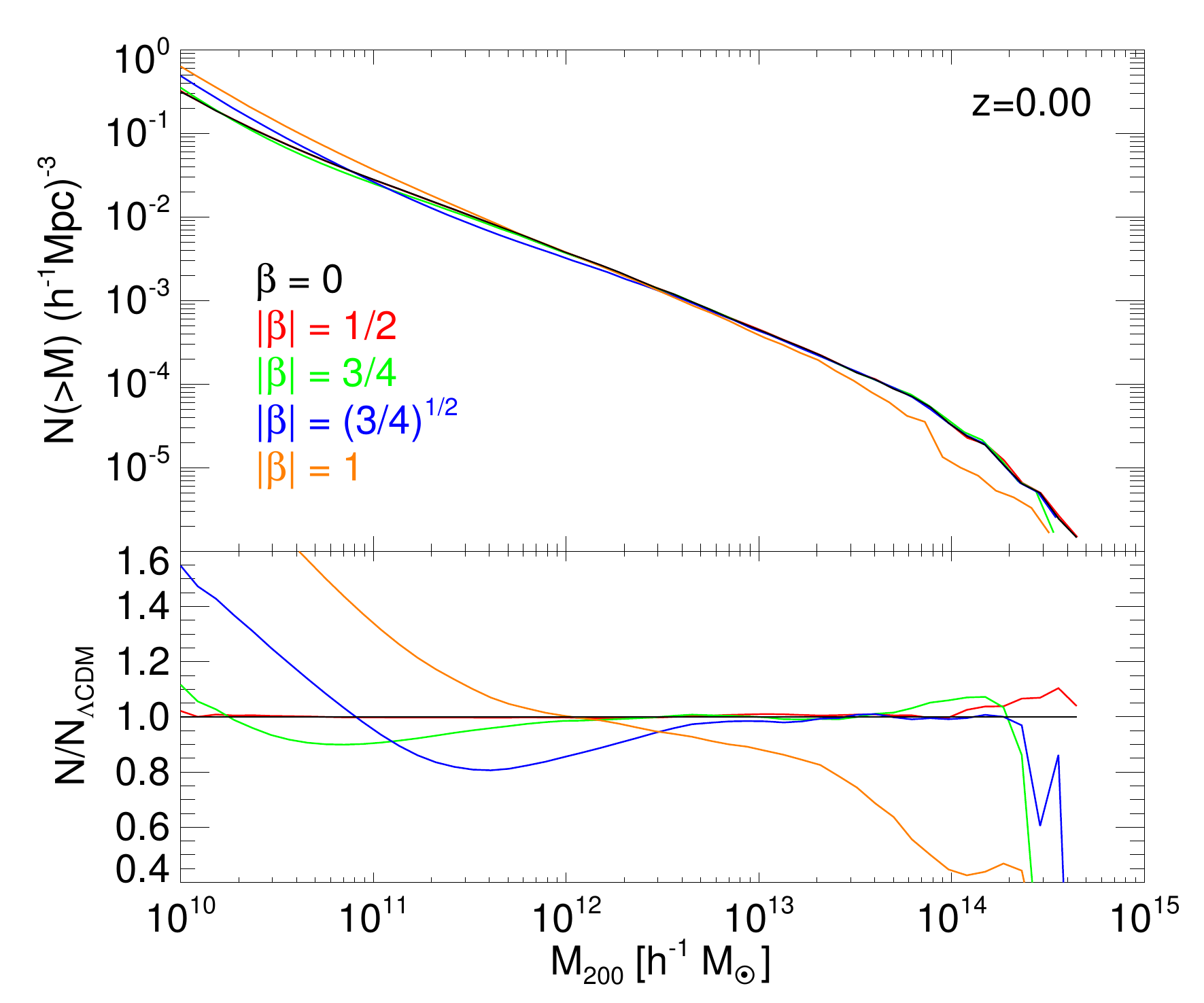}
\end{center}
\caption{The cumulative halo mass function at different redshifts for the various McDE scenarios considered in the present work. Again, to avoid too crowded plots, we do not include in the figure the $|\beta |=\{ 7/10\,, 8/10\}$ cases as they are qualitatively consistent with the rest of the models.}
\label{fig:HMF}
\end{figure*}

We now move to a more quantitative analysis of the statistical properties of collapsed structures in McDE cosmologies by computing the abundance of halos as a function of their mass, i.e. the halo mass function (HMF), for the different McDE models under investigation. To avoid any possible impact of the coupling on the identification of gravitationally-bound structures we will rely on the FoF halo catalogue to compute the HMF of all the models.
In Fig.~\ref{fig:HMF} we show the cumulative FoF HMF for the models $|\beta |=\{ 0\,, 1/2\,, 3/4\,, \sqrt{3}/2\,, 1\}$ at five different redshifts $z=\{0\,, 0.5\,, 1\,, 1.5\,, 2\}$. The remaining models $|\beta |=\{7/10\,, 8/10\}$ show a consistent behaviour with the plotted results and have been omitted to avoid crowded figures.\\

The most prominent feature appearing in the HMFs of Fig.~\ref{fig:HMF} is the characteristic shape of the ratio of the halo abundance over the uncoupled case $|\beta |=0$,
showing a significant enhancement at the smallest masses of our sample and a corresponding suppression at intermediate masses. The transition mass between enhanced and suppressed halo abundance moves towards larger masses at progressively lower redshifts, and such evolution appears to be more significant for larger coupling values. Such
result is clearly consistent with the already discussed phenomenon of halo fragmentation, with intermediate mass halos splitting into smaller objects of roughly half the original mass. 
Therefore, in McDE cosmologies the universe is progressively depleted of intermediate-mass halos and correspondingly filled with new halos of smaller mass resulting from the fragmentation of the former.
As the halo fragmentation is triggered by the differential friction terms on the two CDM particle species, it starts from low-mass halos as these are the first to form and have higher peculiar velocities, and then progressively involves halos with larger masses. Also, such process evolves faster for larger values of the coupling. This is particularly evident by comparing the HMF of the $|\beta |=\sqrt{3}/2$ and $|\beta |=1$ cosmologies, with the latter showing a much faster evolution of the HMF ratio and 
a shift of the transition mass between low-mass enhanced abundance and high-mass suppressed abundance of more than one order of magnitude between $z=2$ and $z=0$. 

It is also interesting to notice how the unitary coupling $|\beta | = 1$ predicts a significant suppression of the high-mass tail of the HMF at all redshifts, becoming more dramatic for $z\lesssim 0.5$. Such feature might be appealing in relation to the presently debated issue of the reported tension between the latest cosmological constraints and the abundance of detected SZ clusters from the Planck satellite mission \citep[see][]{Planck_XX,Planck_016}. In fact, the expected number of clusters appears significantly reduced 
in McDE models with sufficiently large couplings,
without affecting the normalisation of linear perturbations at large scales, i.e. without changing $\sigma _{8}$. 
In other words, the halo fragmentation process is found to suppress the growth of cluster-sized halos at low redshifts, and to significantly reduce the expected number of clusters as compared to $\Lambda $CDM for a given $\sigma _{8}$ normalisation, thereby alleviating the apparent tension between the latest Planck data and the predictions of a standard $\Lambda $CDM evolution of cosmic structures.

However, at the same time the strongly increased abundance of small mass halos is likely to be in tension with astrophysical data at small scales, as it exacerbates the longstanding ``missing satellite" problem of CDM hierarchical structure formation. The latter effect therefore promises to be one of the most constraining predictions of McDE scenarios in the nonlinear regime, possibly allowing to put firm constraints on the McDE interaction strength below the unitary coupling value that seems to be required to address the unexpectedly-low observed abundance of galaxy clusters by the Planck satellite. It is nonetheless interesting to notice how McDE cosmologies can suppress the abundance of massive clusters for a fixed $\sigma _{8}$ linear amplitude, as such behaviour can hardly be obtained with any other DE or MG model once the background expansion is kept fixed.

\subsection{Structural properties of CDM halos}
\begin{figure*}
\begin{center}
\includegraphics[scale=0.29, angle=90]{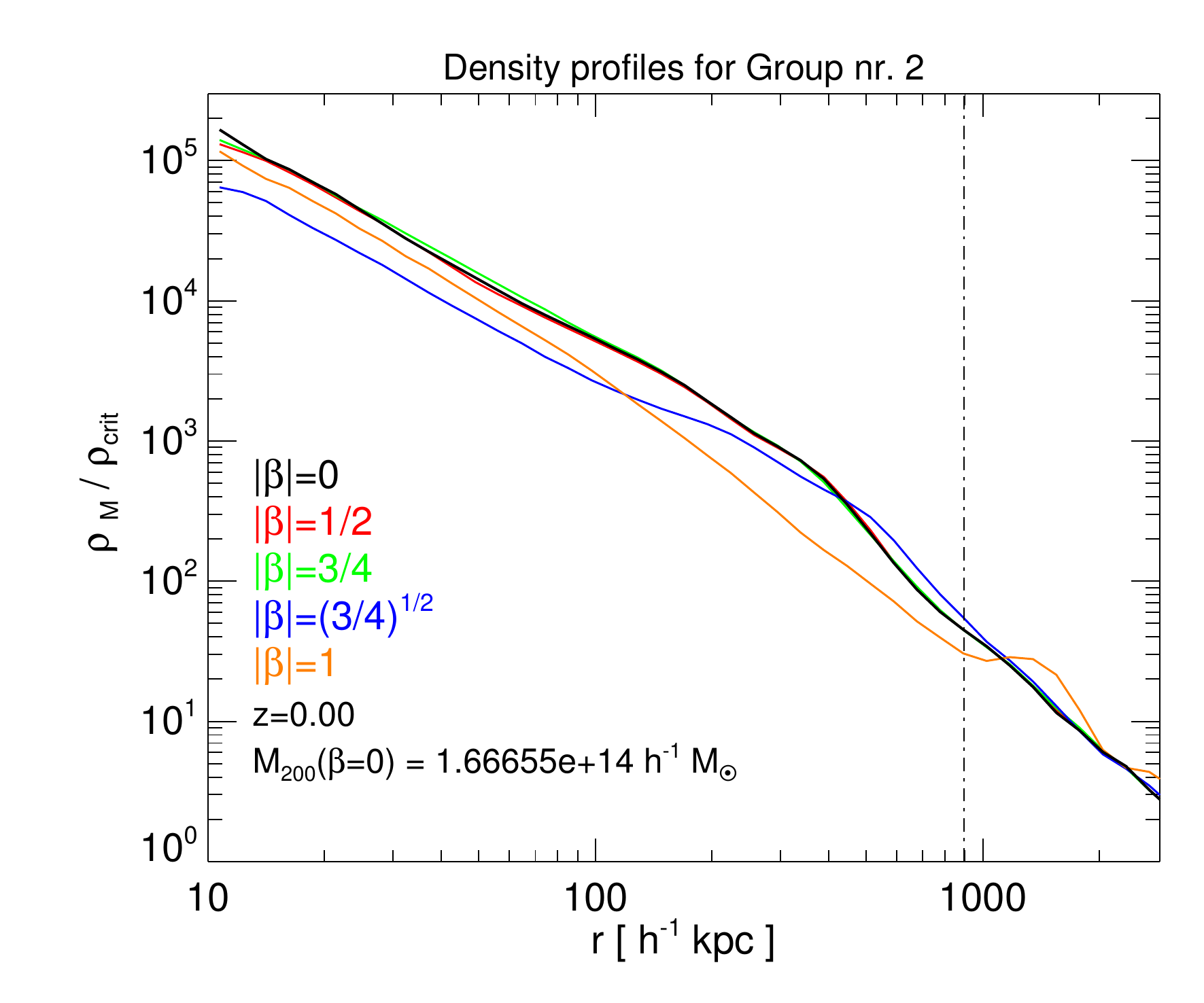}
\includegraphics[scale=0.29, angle=90]{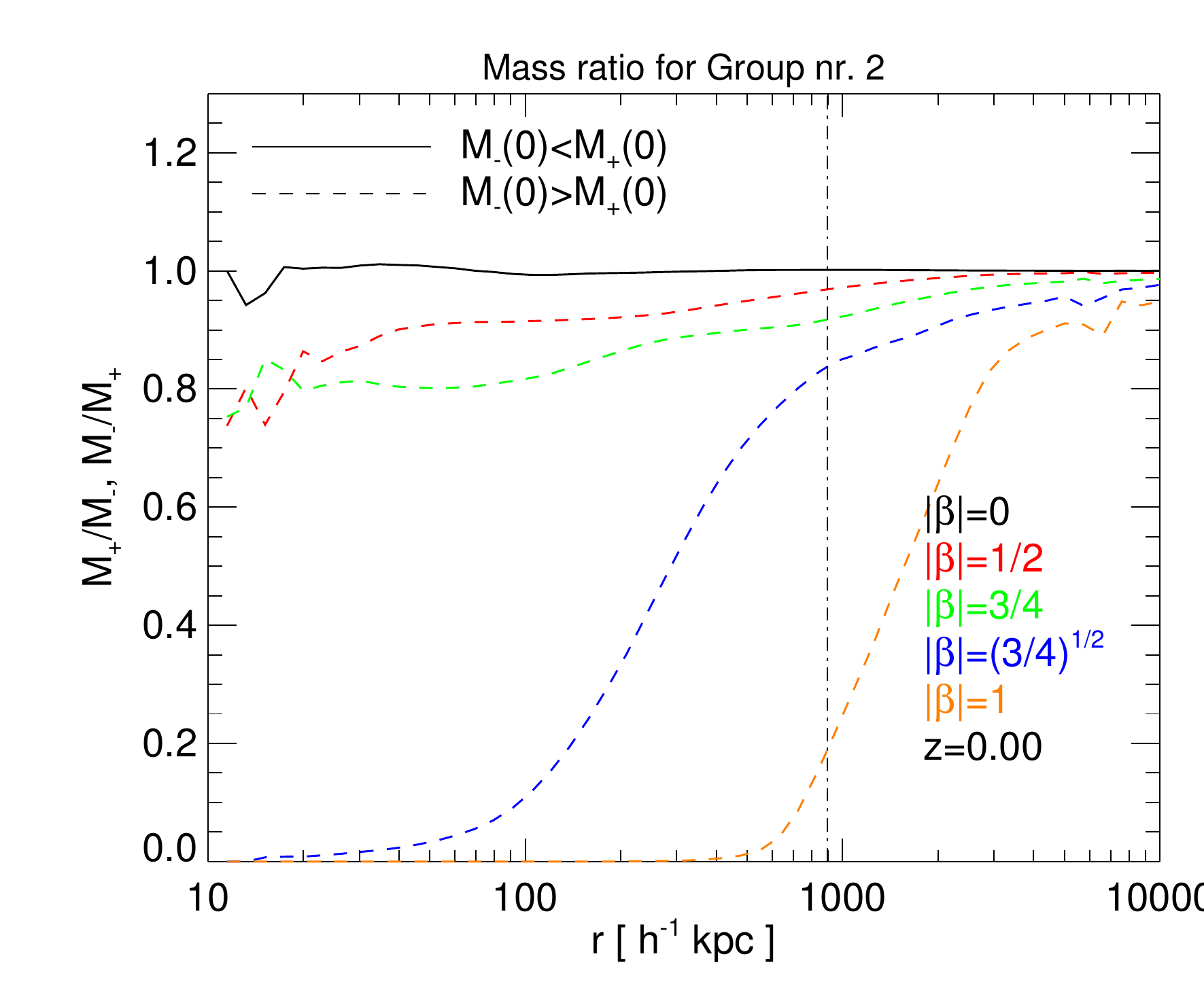}
\includegraphics[scale=0.29, angle=90]{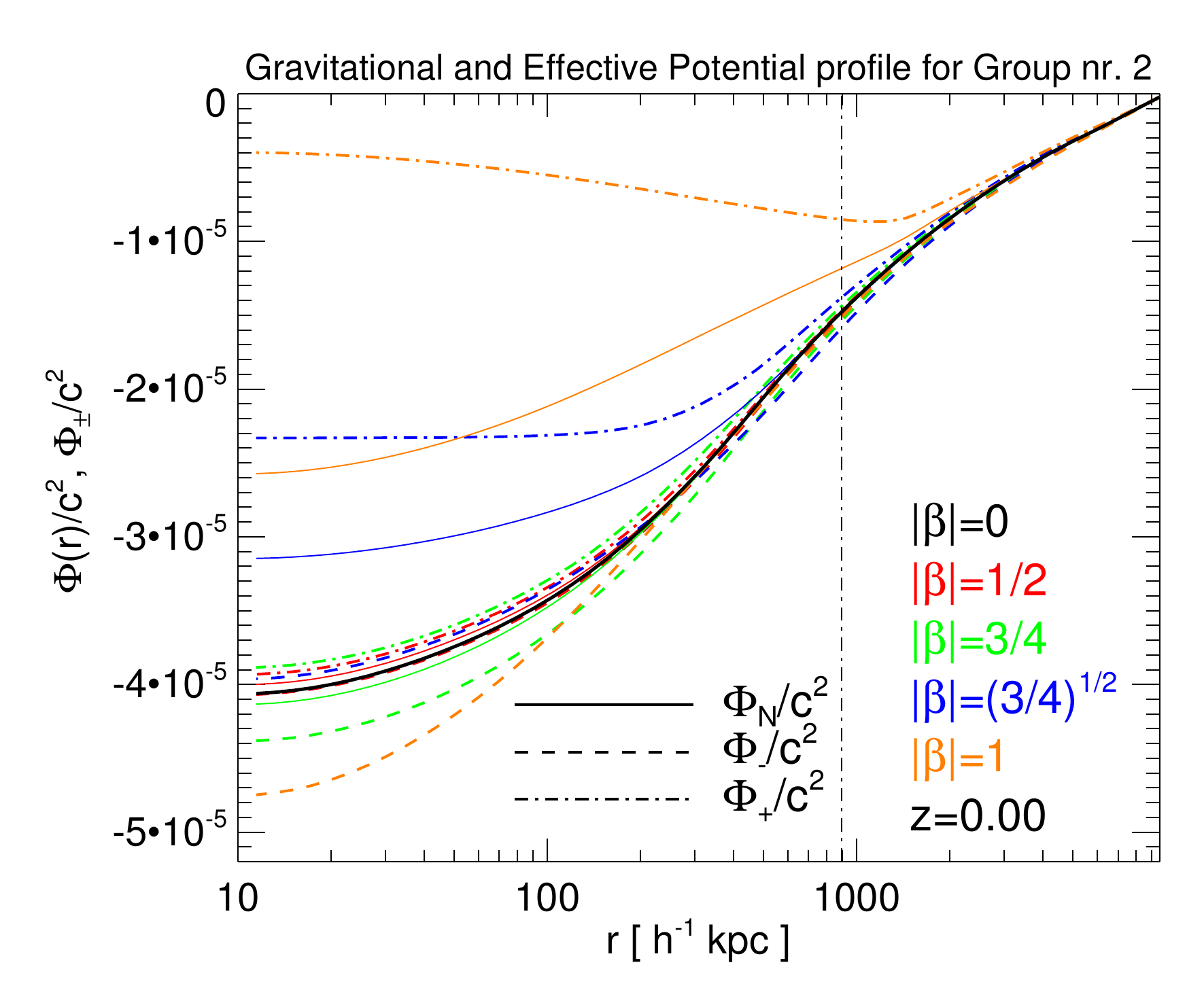}\\
\includegraphics[scale=0.29, angle=90]{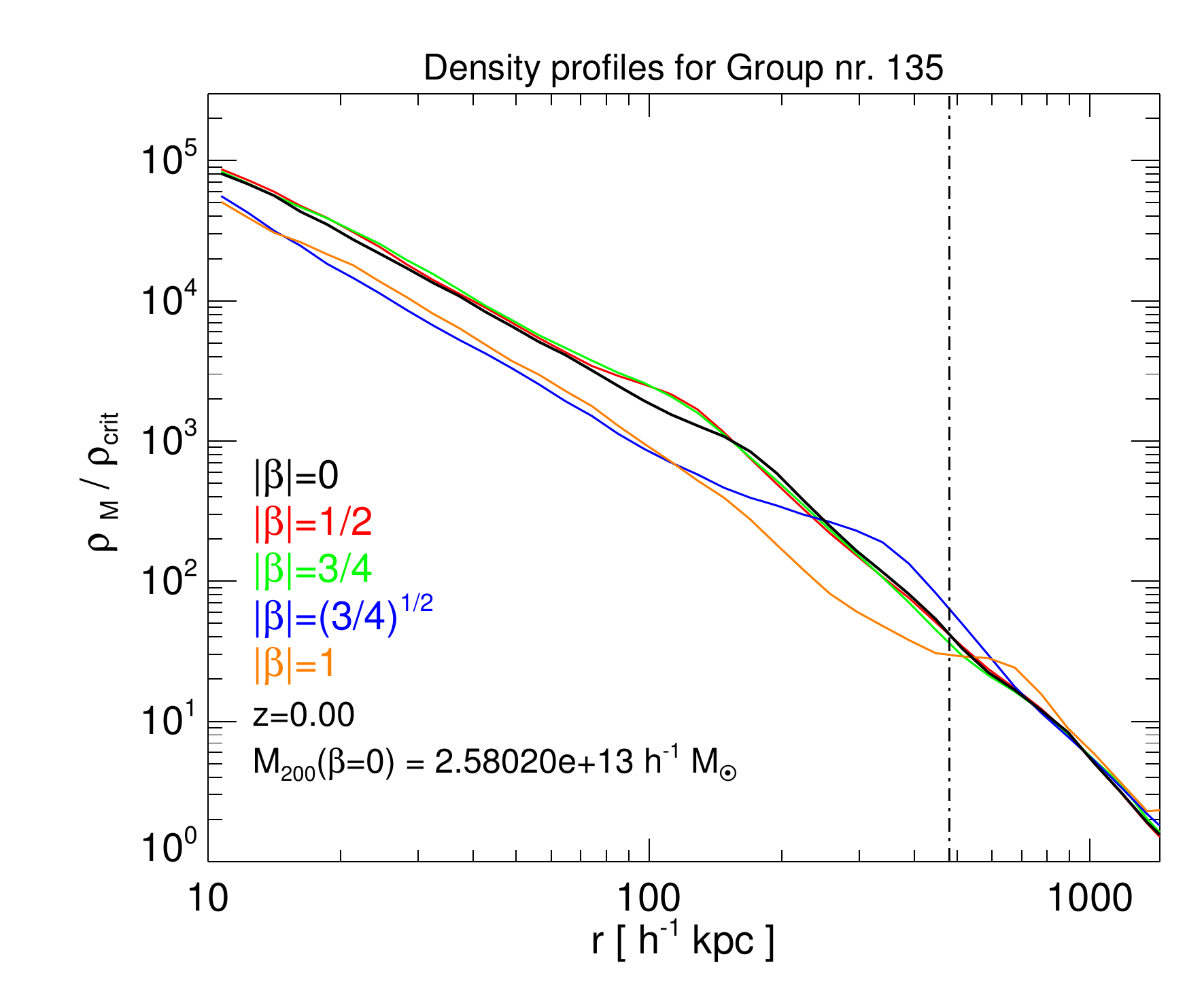}
\includegraphics[scale=0.29, angle=90]{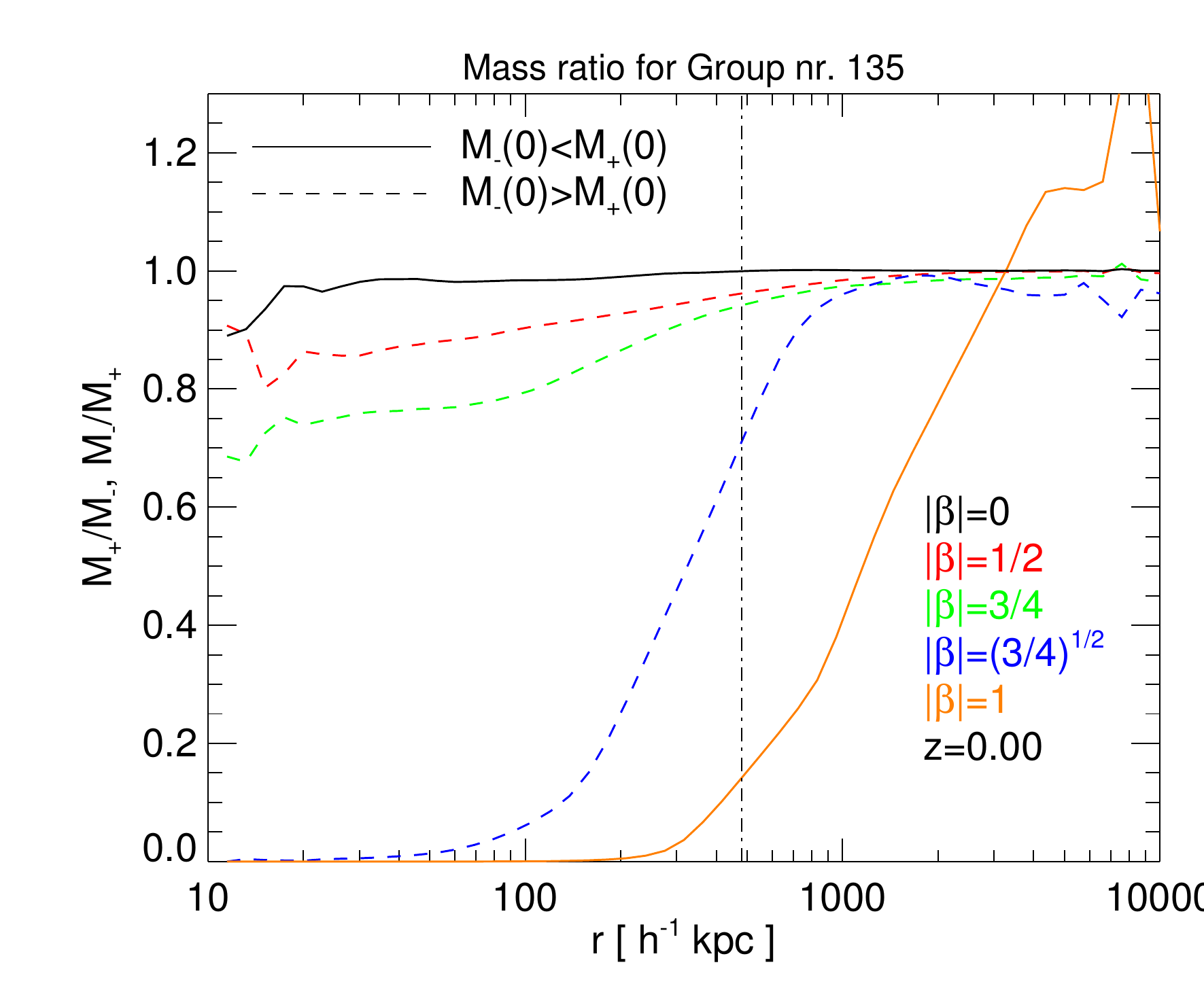}
\includegraphics[scale=0.29, angle=90]{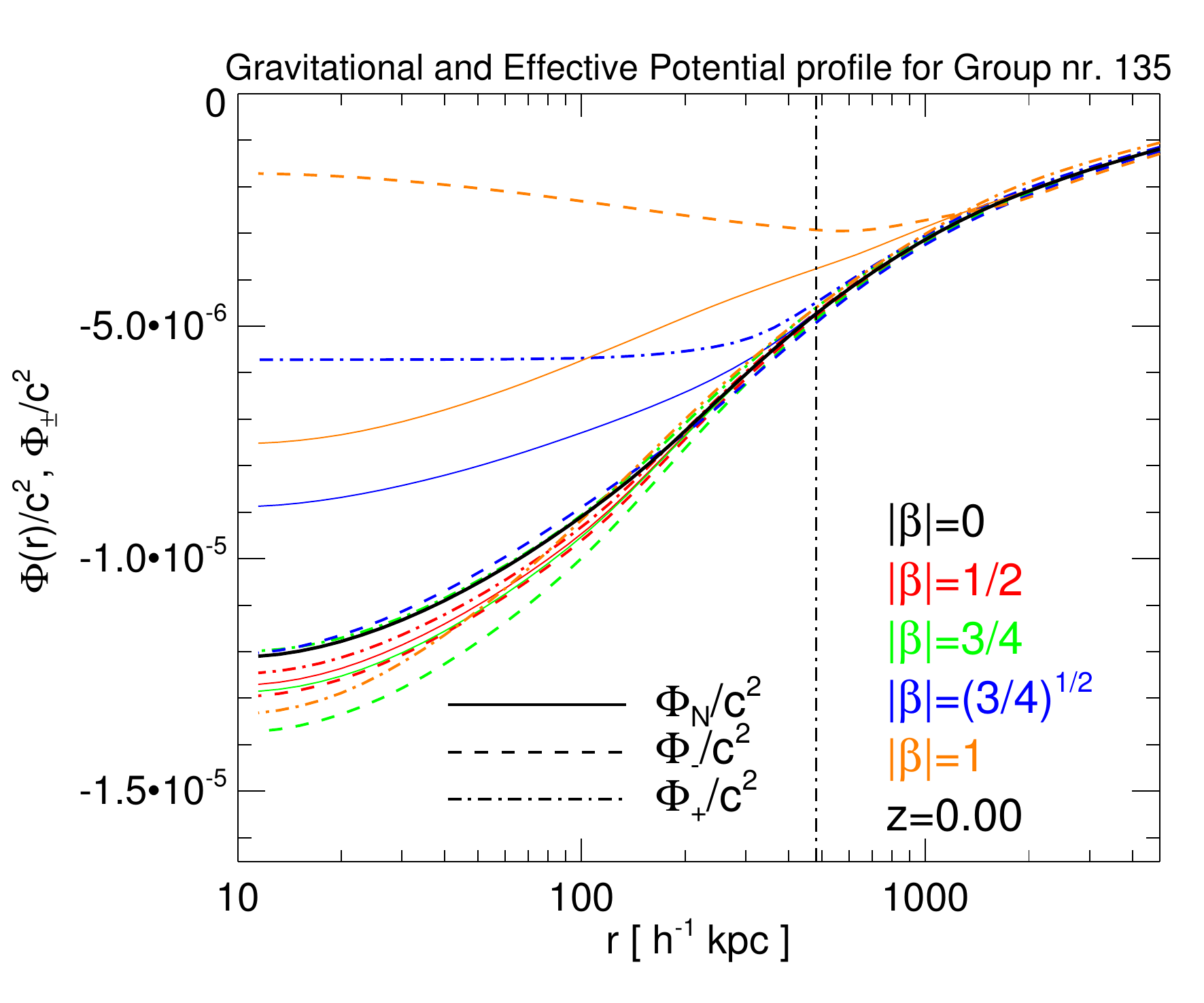}\\
\includegraphics[scale=0.29, angle=90]{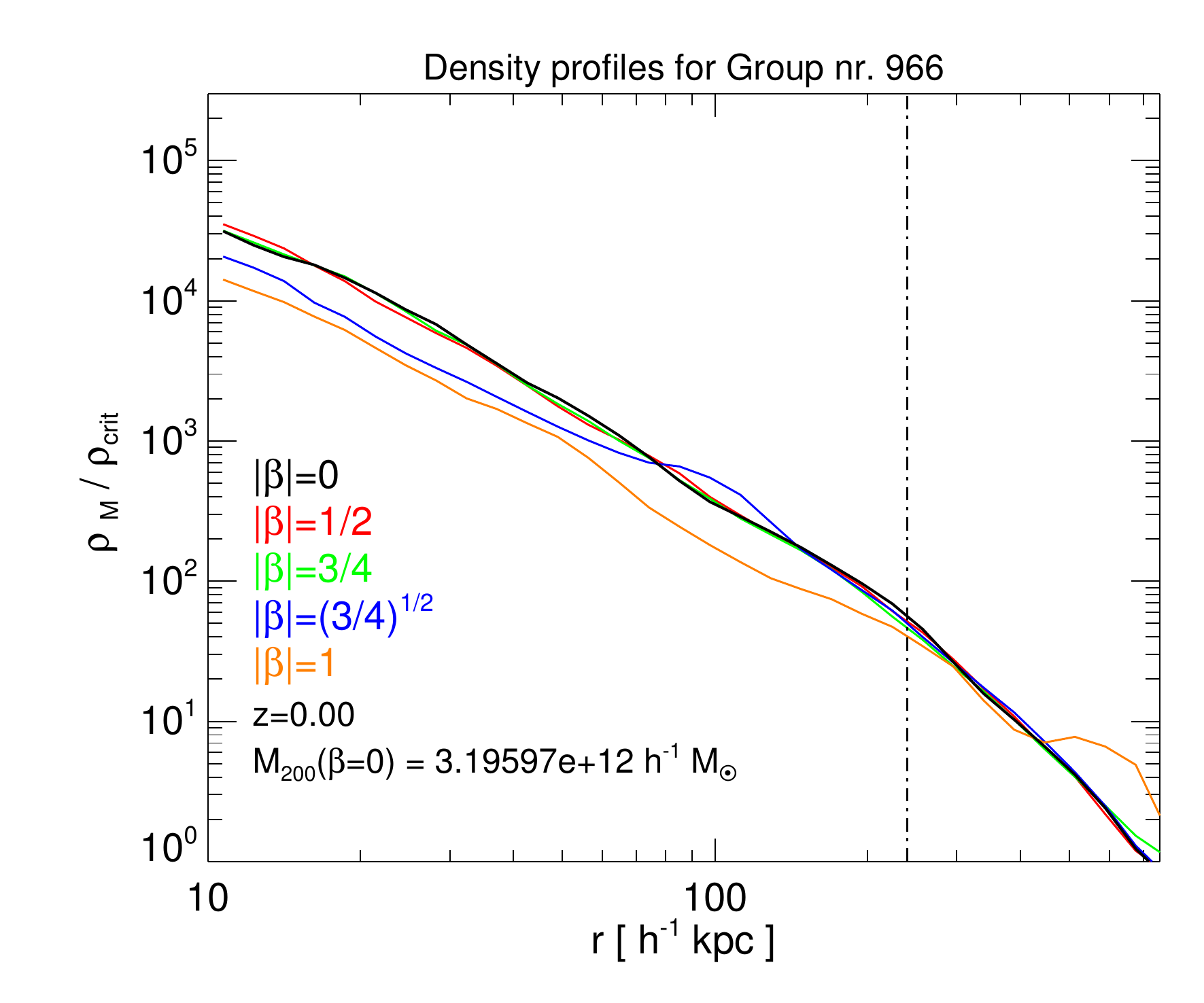}
\includegraphics[scale=0.29, angle=90]{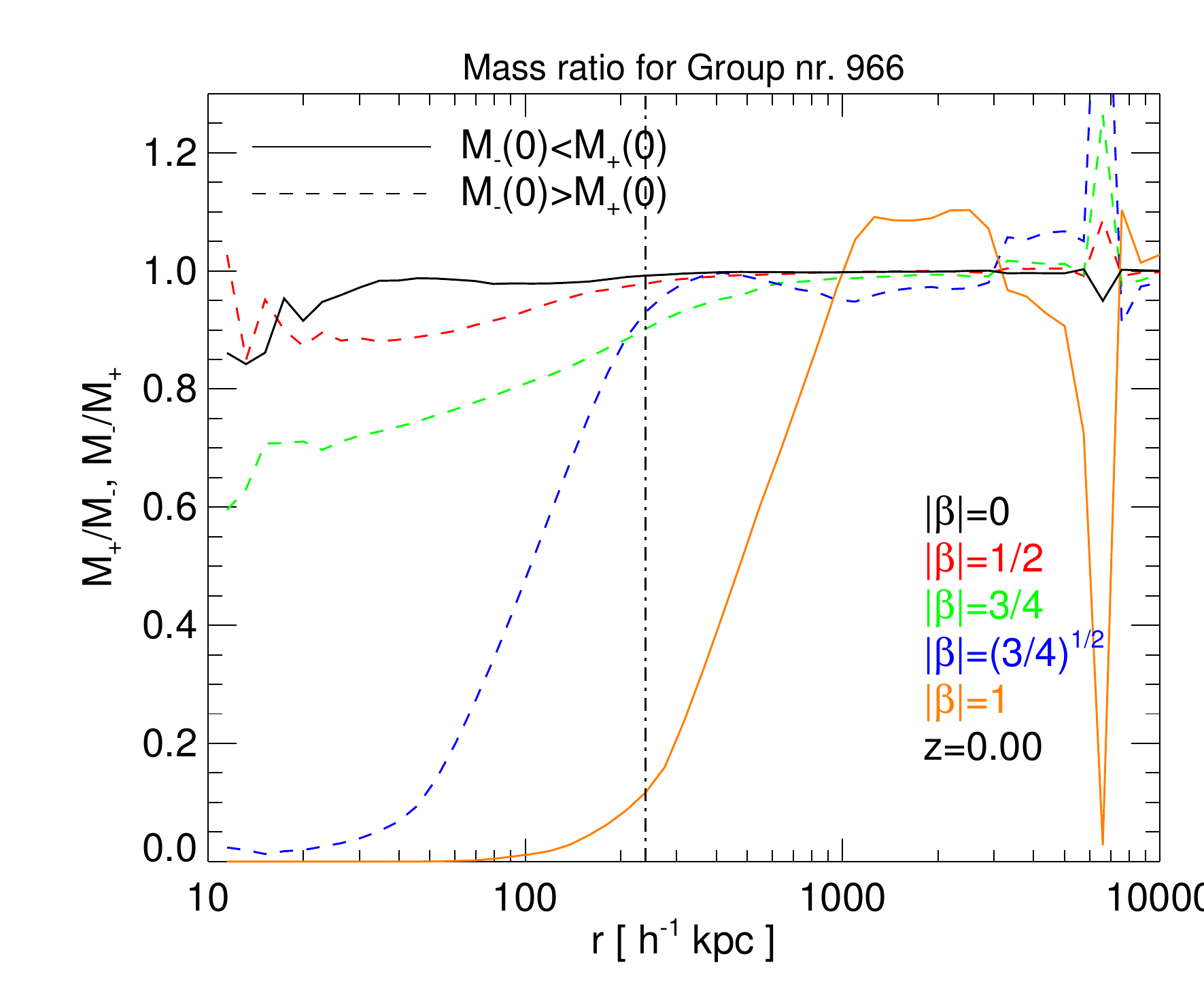}
\includegraphics[scale=0.29, angle=90]{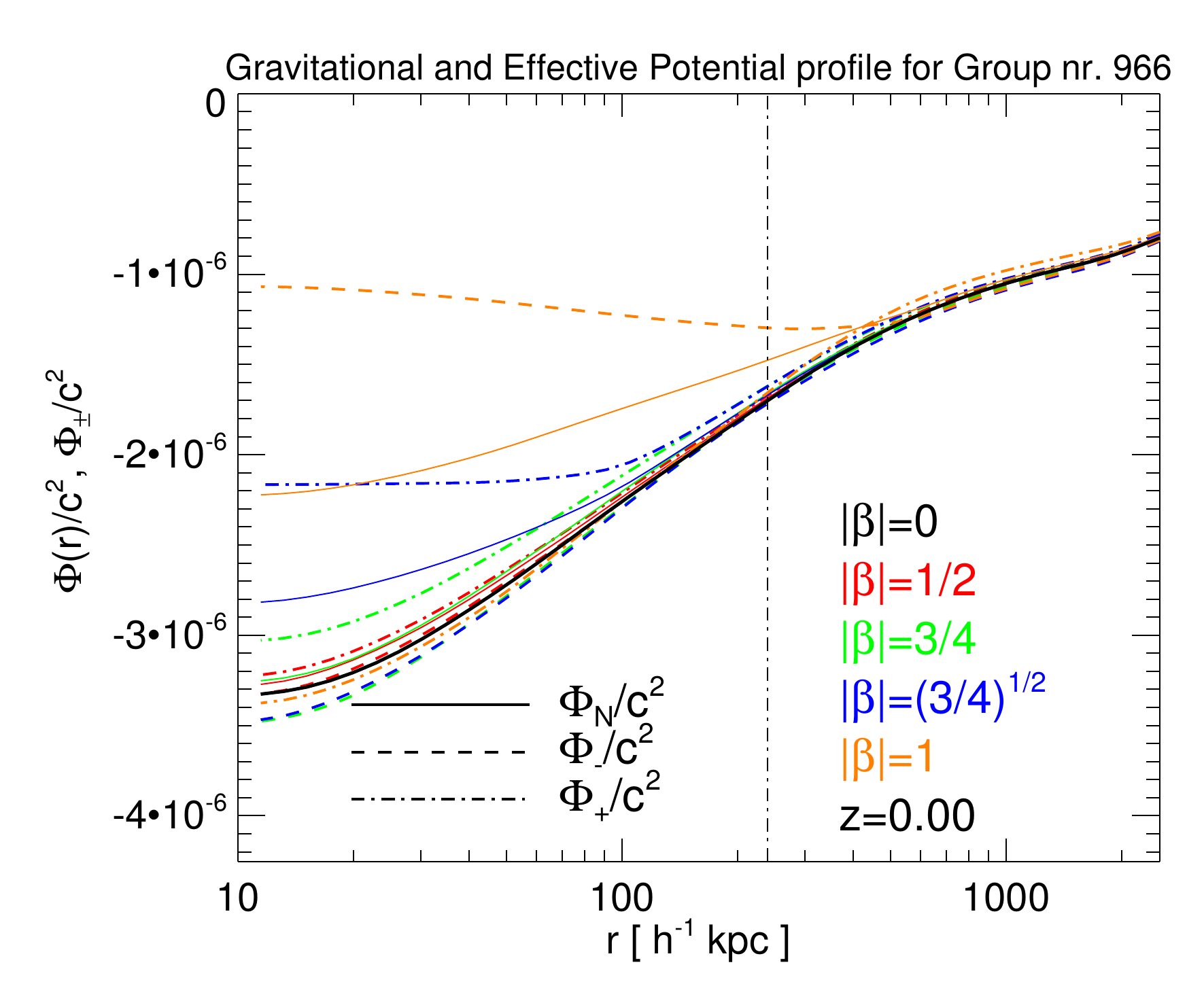}
\end{center}
\caption{The density profile ({\em left}), mass ratio profile ({\em middle}), and gravitational potential profile ({\em right}) for three halos of different mass within of our matched catalog.}
\label{fig:structural_properties}
\end{figure*}

We now move to the highly nonlinear regime of McDE models by investigating the structural properties of CDM halos at $z=0$. 
In the present work we will focus mainly on the matter distribution within the halos and on the consequent space-dependent screening
of the scalar fifth-forces, leaving a more extended analysis of other structural and dynamical properties (such as e.g. halo ellipticity, spin, velocity dispersion, etc...) to future work.\\

Since the same seed was adopted for
the random realization of the total matter power spectrum in the initial conditions
of all the different runs, structures are expected to form
roughly at the same positions in all simulations, making it
possible to identify the same objects in all the simulations and to
compare their properties on an individual basis. However, for increasing values of the coupling the small-scale dynamics
of CDM particles shows progressively stronger deviations from the uncoupled case, up to the onset of 
the halo fragmentation processes widely discussed above, thereby introducing a coupling-dependent
offset of corresponding objects in the different runs, and eventually (for large enough coupling values, typically $|\beta |\gtrsim \sqrt{3}/2$)
determining a complete mismatch of the individual structures at small scales from the uncoupled cosmology.
In order to allow a safe comparison of the inner properties of CDM halos in the different cosmologies, we therefore apply
a selection criterion -- already described in \citet{Baldi_etal_2010} -- and identify objects found in the different
simulations as being the same structure only if the centre 
of each of them lies within the virial radius of the corresponding
structure in the $|\beta |=0$ run. If this criterion is not fulfilled
for all the different simulations we want to compare, we just do not
include the corresponding halo in any of the direct comparisons
described below. Clearly, such matching procedure will produce
a catalog with a variable number of matched objects depending on the range of models that is included in the comparison.
If not stated otherwise, in the following we will restrict such range to the same 5 models considered
in Fig.~\ref{fig:HMF} and we will apply our procedure to the 1000
most massive haloes identified by the FoF algorithm in the different runs. The inclusion of the gravitational coupling $|\beta |=\sqrt{3}/2$ and of the unitary coupling model $|\beta |=1$, for which the halo fragmentation process occurs, as expected strongly reduces the number of objects that fulfill
our matching criteria, such that the full matched catalog contains only 24 objects out of the 1000 FoF halos initially considered.
For all such 24 objects we have computed the spherically averaged density profiles around the most bound particle identified with the {\small SUBFIND} algorithm,
the spherically averaged density ratio of the two CDM particle species, and the radial profile of both the standard Newtonian potential $\Phi _{N}(r)$ and the total potentials $\Phi _{\pm}(r)$, defined as the potentials experienced by a positively- or negatively-coupled test particle, respectively. 
In Fig.~\ref{fig:structural_properties} we show such comparisons for 3 randomly chosen halos of different total virial mass, namely  $M_{200}\approx 10^{12}$, $10^{13}$, and $10^{14}$ M$_{\odot }/h$. 

More specifically, the left column of plots of Fig.~\ref{fig:structural_properties} displays the density profile of the three randomly chosen halos in the different
McDE models, and clearly shows the impact that the different coupling values have on the total CDM distribution, in particular in the innermost regions of the halos. As one can see in the figures, couplings as large as $|\beta |=3/4$ appear to have a very little impact on the total density profile of CDM halos over a wide range of halo masses. On the contrary, for the gravitational and unitary couplings $|\beta |=\sqrt{3}/2$ and $|\beta |=1$ the impact of the DE interactions starts to be clearly sizeable, with both models resulting in a lower CDM overdensity in the inner parts of the halos as compared to the uncoupled case. Furthermore, both models are characterised by a clear bump in the halo density profiles at larger radii. Such bump is produced by the outflow of one of the two CDM species under the effects of its friction term and repulsive fifth-force. As expected, the bump appears at larger radii for the largest coupling value, consistently with an earlier onset of halo fragmentation and with a faster decoupling of the two CDM particle distributions due to a stronger fifth-force repulsion.

Such dynamical decoupling of the two different CDM species within individual structures can be better understood by looking
at the middle column of Fig.~\ref{fig:structural_properties}, where the radial profile of the ratio between the density of the dominant and the sub-dominant CDM species is shown for the same McDE cosmologies. In the plots we display  the density ratio with solid lines if the halo core
is dominated by the positively-coupled species, while dashed lines are used in the opposite case. As one can see from the figures, the latter situation is the most likely for coupling values up to the gravitational coupling
(for 22 out of 24 objects of our matched sample the core is dominated by negatively-coupled particles in the $|\beta |\leq \sqrt{3}/2$)
while for the unitary coupling $|\beta |=1$ the occurrence of positively-coupled particles dominating the core is more frequent (8 cases out of 24).
The plots show that a non-negligible imbalance (at the level of $\approx 10-20\%$) between the two CDM particle types is present in the core of CDM halos even
for coupling values of $|\beta |=1/2$ and $|\beta |=3/4$, while this becomes much more dramatic for $|\beta |\geq \sqrt{3}/2$, with the volume
within the virial radius of the structure being fully dominated by only one CDM particle species. This behaviour shows how the halo fragmentation process works
at highly nonlinear scales, with the particles of one of the two CDM species outflowing the halo under the effect of their different friction term and 
of the repulsive interaction with the dominating particle type, accumulating in the outskirts of the structures and then eventually recollapsing together under their strongly enhanced gravitational attraction (by the attractive fifth-force acting between particles of the same type) to form a separate object. The accumulation of particles in the outskirts of halos is even directly visible  for some of the objects in the unitary coupling case as a ``bump" of the density ratio just out of the halo virial radius.

Finally, in the right column of Fig.~\ref{fig:structural_properties} we display, for the same three randomly chosen halos, the spherically-averaged profile
of both the standard Newtonian gravitational potential $\Phi _{\rm N}(r)$ (as solid lines) and of the two total potentials $\Phi _{\pm }(r)$
(as dashed and dot-dashed lines) resulting from the sum of the standard gravitational potential and of the attractive and repulsive fifth-forces determined by the radial distribution of the two CDM particle species. In other words, the total potentials $\Phi _{\pm }(r)$ represent the real potential experienced by a 
positivley- ($+$) and negatively- ($-$) coupled test particle located at a distance $r$ from the halo centre. Very interestingly, we can immediately notice how in the McDE models the standard potential wells of CDM halos are in general shallower than in the uncoupled case, as a consequence
of the lower overdensity  in the inner regions of the halos, with the effect becoming particularly significant for $|\beta |\geq \sqrt{3}/2$. 
It is also very interesting to notice how the total potentials $\Phi _{\pm }(r)$ have very different radial shapes from the standard Newtonian potential and
from one another, and how their relative behaviour is correlated with which of the two CDM species dominates the halo core. In particular,
one can notice (more evidently when looking at the curves associated with the largest coupling values) how the total potential experienced
by the dominant particle species is in general steeper than the standard Newtonian potential arising in the same model, thereby favouring a further concentration of the dominant particles in the halo core. 
This is clearly due to the attractive fifth-force that acts on the particles in addition to the standard Newtonian attraction.
On the contrary, the total potential experienced by the sub dominant particle species is much shallower than
the Newtonian potential and might show a very extended plateau (see e.g. the blue curves for the $|\beta |=\sqrt{3}/2$ case) or even a negative
radial derivative, meaning that sub-dominant particles are repelled away from the halo core (as for the case of the unitary coupling). In the latter case, the total potential experienced by the sub-dominant particles shows a minimum just outside the halo virial radius, thereby favouring
the accumulation of particles in a spherical shell around the halo, consistently with what shown by the radial profile of the CDM particle species density ratio.
Also in this case, such behaviour is the result of the fifth-force correction (this time repulsive) to the standard Newtonian interaction.\\

These results show for the first time how the effective suppression of the coupling in McDE models extends also to the highly nonlinear regime of structure formation, providing a new type of screening mechanism of a scalar fifth-force around collapsed structures. Differently from other
types of screening mechanisms (as e.g. the chameleon, symmetron, or Vainshtein mechanisms) where the fifth-force is simply suppressed in overdense regions of the universe, in McDE cosmologies the screening is provided by the balance between attractive and repulsive corrections to standard gravity and depends on the local distribution of the two different CDM particle species. As a result, the fifth-force is screened at a different level in different structures and for different particle types, resulting in a segregation of the two CDM particle species and in a direct violation of the weak equivalence principle. Furthermore, the screening appears as
an unstable phenomenon since the halo fragmentation process described above does progressively split halos of mixed particle species into
pairs of halos dominated by one single particle type, for which the screening mechanism is obviously absent. Therefore, the McDE screening mechanism appears as a transient phenomenon characterising halos before fragmentation, and involving structures of different mass at different redshifts since the halo fragmentation proceeds in a hierarchical fashion starting from the smallest objects (as shown in Fig.~\ref{fig:type_fraction}).\\

\begin{figure*}
\includegraphics[width=1.7in]{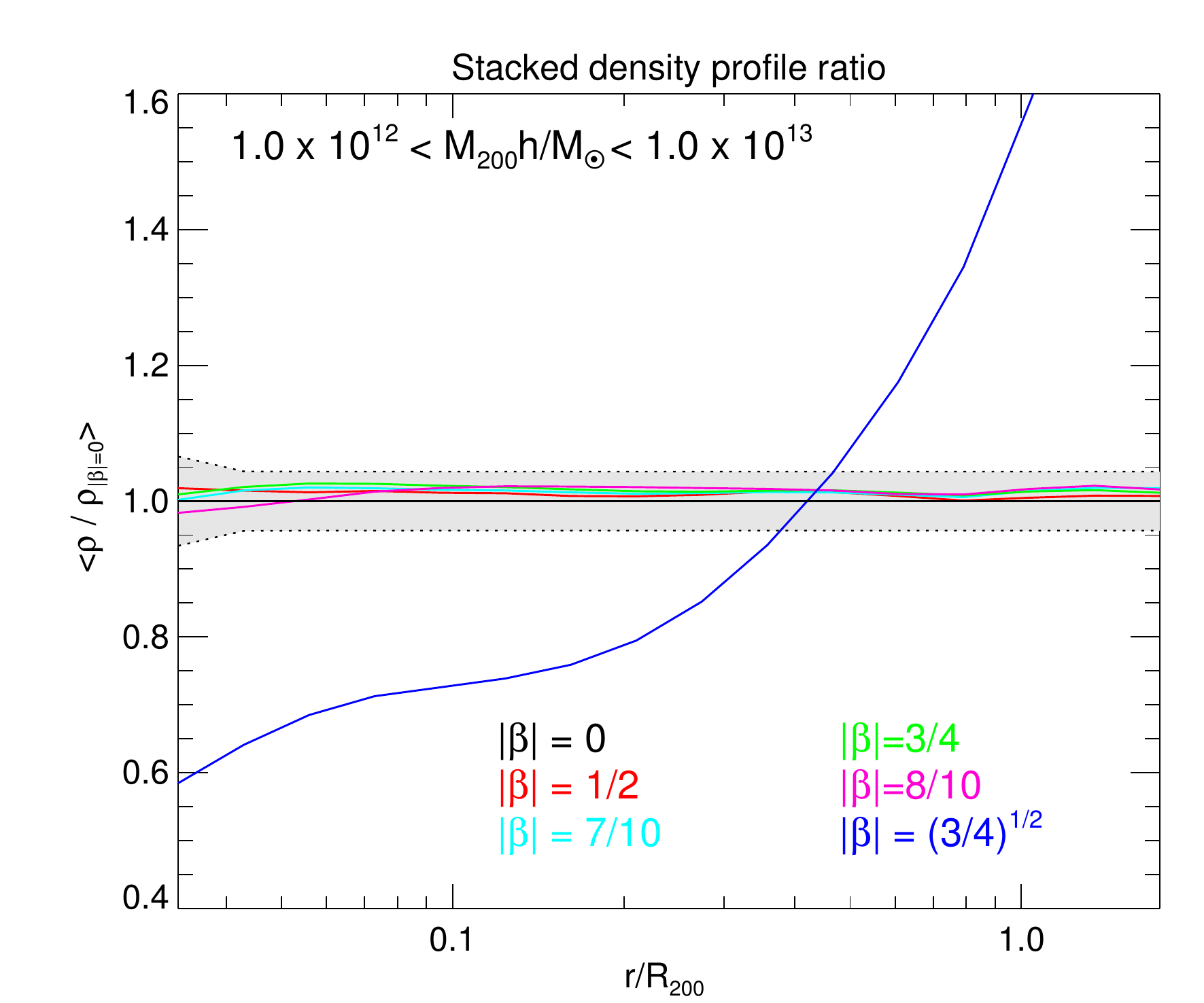}
\includegraphics[width=1.7in]{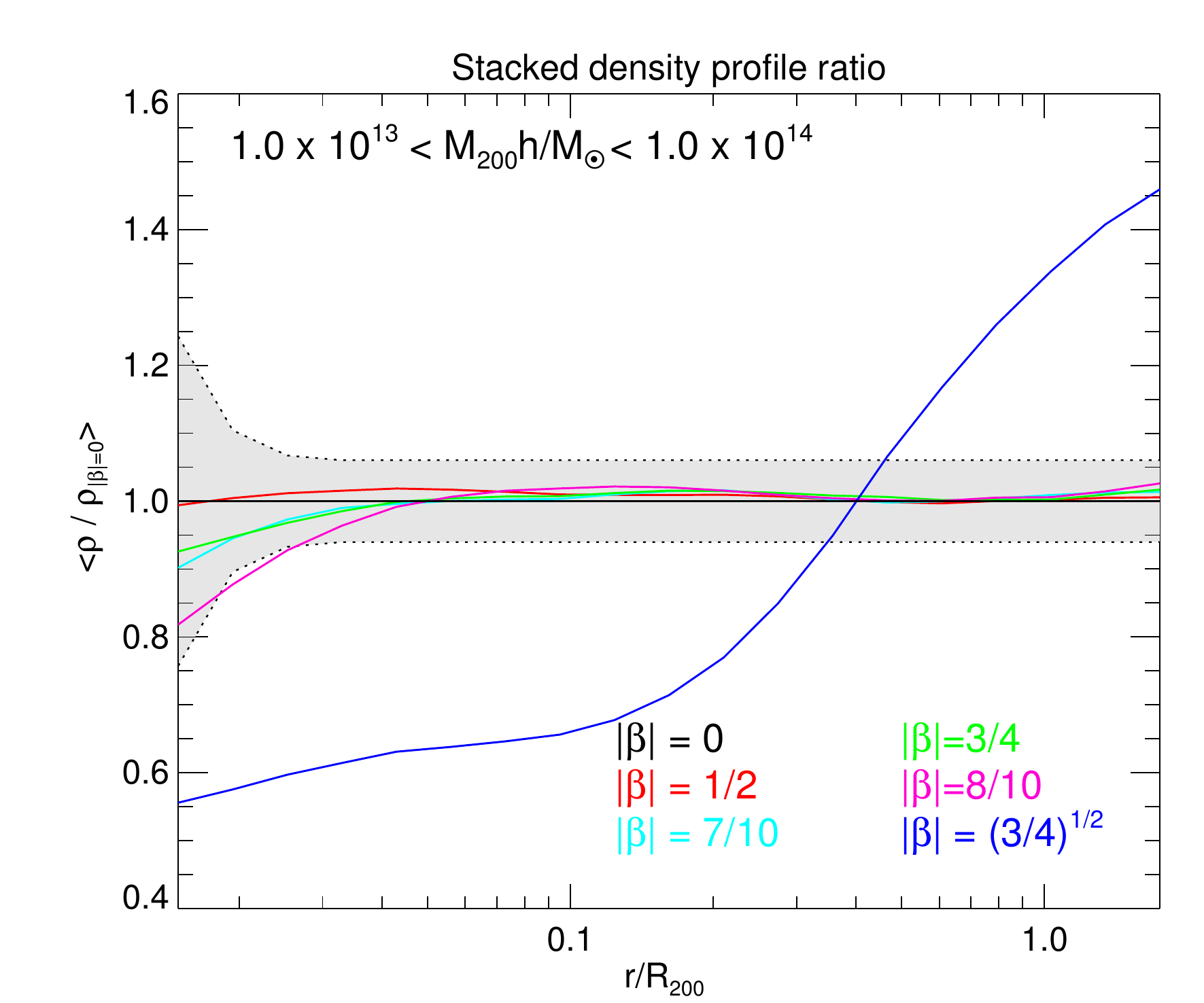}
\includegraphics[width=1.7in]{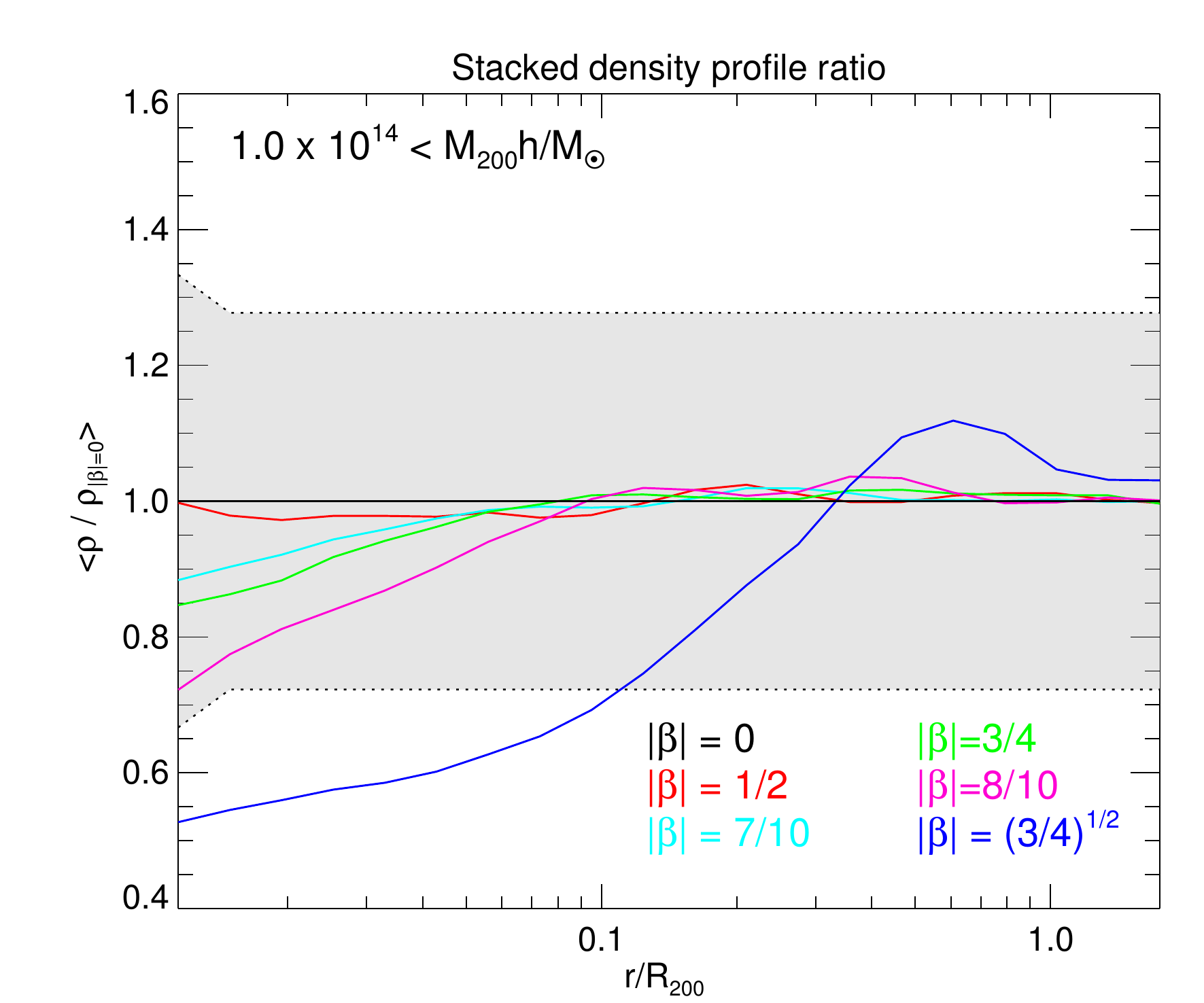}
\caption{Stacked density profile ratio}
\label{fig:profile_ratio}
\end{figure*}

To conclude our analysis of the structural properties of CDM halos in McDE models, we investigate the stacked density profiles of halos in three
different mass ranges over the sample of the 1000 most massive halos identified in each simulation. In order to allow a statistically significant comparison 
we exclude the unitary coupling from the analysis and apply again the matching procedure described above to the remaining 6 models of our
sample $|\beta |=\{ 0\,, 1/2\,, 7/10\,, 3/4\,, 8/10\,, \sqrt{3}/2\}$, as this results in a much larger sample of matched structures. For all such structures,
we compute the density profile as a function of the effective radius defined as the physical radius in units of the virial radius, $r/R_{200}$, and 
plot the averaged density profile ratio to the uncoupled case over 20 radial bins. The result of such procedure is displayed in Fig.~\ref{fig:profile_ratio}
for halos in the mass ranges $10^{12}\leq M_{200}h/M_{\odot} < 10^{13}$ ({\em left}), $10^{13}\leq M_{200}h/M_{\odot} < 10^{14}$ ({\em middle}), and $M_{200}h/M_{\odot} \geq 10^{14}$ ({\em right}), where the shaded areas represent the Poissonian error on the averaged density profile ratios.
As on can see from the plots, in all the three different mass ranges, all McDE models except the gravitational coupling $|\beta |=\sqrt{3}/2$ appear to be
consistent with the standard Navarro-Frenk-White \citep[NFW][]{NFW} density profile of the uncoupled case within statistical errors, even though a coherent trend of density suppression in the halo core clearly appears for the most massive objects. Such result is consistent with the observation discussed above that for $|\beta |<\sqrt{3}/2$ CDM halos do not fragment into smaller and single-particle-type dominated objects, which preserves the efficiency of the McDE screening
of the scalar fifth-forces, such that no major effects on the internal structure of halos is observed. On the other hand, the $|\beta |=\sqrt{3}/2$ model shows a very pronounced feature in the averaged density profiles, characterised by a strong suppression of the overdensity in the inner regions, and by a correspondingly strong enhancement of the density contrast in the outskirts of the halos. Such effect shows also a very clear evolution with halo mass, being more pronounced for small mass objects than for cluster-sized halos. This is also consistent with the observation of the hierarchical nature of the halo fragmentation process that starts from the smallest objects and then proceeds to larger structures.\\

As a general result of our investigation of structural properties of CDM halos in McDE cosmologies, we can therefore conclude that for models with
couplings below the gravitational strength $|\beta |< \sqrt{3}/2$ the scalar interactions have a very mild impact also on the nonlinear dynamics of CDM particles at small scales, besides leaving completely unaffected the background and linear perturbations evolution.  For such models the only potentially observable effects 
that we identified through our analysis
are related to a $\approx 10-20\%$ reduction of the number of low mass halos ($10^{10}-10^{11}$ M$_{\odot }/h$) possibly associated to a corresponding increase of the number of objects at even smaller masses (which are below the resolution available to our simulations), and to a mild suppression of the concentration of cluster-sized CDM halos. On the other hand, as soon as the gravitational strength $|\beta |=\sqrt{3}/2$ is reached and overcome, the onset of halo fragmentation breaks the
screening of the fifth-forces, and McDE cosmologies show very significant effects on the abundance of halos and on their internal structures, including their density profiles and the associated gravitational potentials.\\

\section{Conclusions}
\label{sec:concl}

In this paper, we have presented the outcomes of the first high-resolution N-body simulations of Multi-coupeld Dark Energy cosmologies. These
are cosmological models characterised by the existence of two distinct species of CDM particles with opposite couplings to a classical Dark Energy scalar field,
giving rise to both {\em attractive} and {\em repulsive} long-range scalar interactions between CDM particles.
While requiring the same number of parameters as a standard coupled quintessence model, Multi-coupled Dark Energy scenarios
provide a very effective way to screen the coupling during matter domination, thereby strongly alleviating the impact of the interaction between Dark Energy and CDM particles on the background and linear perturbations evolution of the universe. In this respect, Multi-coupled Dark Energy models are 
practically indistinguishable from a standard $\Lambda $CDM cosmology up to very recent epochs for a wide range of couplings. Nonetheless, the effects of the additional fifth-forces in the nonlinear regime of structure formation are expected to imprint characteristic features in the statistical and structural properties of CDM halos that might allow to observationally test the model. 

To this end, a first series of low-resolution N-body simulations was performed by \citet{Baldi_2012b} with the main purpose of sampling the parameter space of the model and highlight such possible observational footprints.
In this work, we did improve with respect to those early simulations in several aspects, by running the first high-resolution simulations of Multi-coupled Dark Energy models that allow to investigate in detail the statistical and structural properties of CDM halos arising in these cosmologies up to a coupling value of $|\beta |=1$. We briefly recap here the main conclusions of our study.\\

First of all, the various effects that we have highlighted with our high-resolution simulations, and that will be listed below, show a very strong dependence on the coupling value, and in particular on whether the coupling is below or above the gravitational coupling $|\beta |=\sqrt{3}/2$. Such value defines the threshold between the regimes where scalar fifth-forces are weaker ($|\beta |< \sqrt{3}/2$) or stronger ($|\beta |> \sqrt{3}/2$) than standard gravity. In general, for all the coupling values of our sample lying in the former range we found very mild effects of the DE-CDM interaction on all the statistical and structural properties of CDM halos that we have investigated; on the contrary, the impact of the interaction becomes very quickly extremely significant as soon as the gravitational coupling threshold is reached and overcome. Therefore, the gravitational coupling $|\beta |=\sqrt{3}/2$ represents an intrinsic threshold for Multi-coupled Dark Energy models, and the small range of coupling values around such threshold promise to determine the most interesting
phenomenological effects of such scenarios, providing at the same time viable and non-trivial observational effects: coupling values much below this threshold appear to be almost indistinguishable from a standard $\Lambda $CDM cosmology, while values significantly above it determine very dramatic effects at nonlinear scales that are likely to be easily ruled out by presently available data. Such effects, according to our present analysis, can be summarised as follows.
\begin{center}
{\em -- Large-scale density smoothing and halo fragmentation -- }
\end{center}
At the largest scales included in our cosmological simulations ($100$ Mpc$/h$ aside) the main effect of the interaction between Dark Energy and CDM in Multi-coupled Dark Energy cosmologies is to smooth the density field of the total CDM fluid, consistently with the earler findings \citep[][]{Baldi_2013} of a significant suppression of power at mildly nonlinear and nonlinear scales. Such effect, however, becomes appreciable only for the largest coupling value included in our sample, $|\beta |=1$, and is almost absent for all the other coupling values considered in our work. At smaller scales, the most prominent effect of the interaction is the fragmentation of bound CDM structures into smaller objects as a consequence of the different dynamical evolution characterising the two distinct CDM particle species. As anticipated above, such phenomenon is completely absent for coupling values below the gravitational coupling threshold $|\beta | < \sqrt{3}/2$, as the gravitational attraction between CDM particles of opposite type is still stronger than their fifth-force repulsion, and sufficient to overcome the effect of the scalar friction that would tend to drag the two different CDM species in opposite directions along their unperturbed trajectory.
However, already for a coupling $|\beta |=\sqrt{3}/2$, our simulations have shown the fragmentation of individual structures into pairs of objects of comparable size, and how the separation of such fragmented halos grows in time along the trajectory of the original parent structure, thereby providing a direct evidence of the violation of the weak equivalence principle in Multi-coupled Dark Energy cosmologies. It is particularly relevant to stress here that even the largest coupling value below the gravitational coupling threshold in our set of models, $|\beta |=8/10$, did not show any sign of halo fragmentation. This gives a feeling of how sharp the transition between the two regimes around the gravitational coupling threshold is.
\begin{center}
{\em -- Halo mass function and suppression of cluster abundance --}
\end{center}
We have computed the abundance of CDM halos as a function of their mass in all the cosmological models under investigation and at different redshifts, based on a halo catalogue computed through a Friends-of-Friends algorithm without any distinction between the two CDM particle species. By comparing the obtained halo mass functions to the uncoupled case, we could highlight a series of characteristic footprints of Multi-coupled Dark Energy scenarios.
Even in this case, such features are either completely absent or extremely weak for coupling values below the gravitational coupling threshold, while becoming very significant at and above the threshold. First of all, as a consequence of the halo fragmentation process there is a clear reduction of the abundance of intermediate mass halos, and a corresponding enhancement of smaller mass halos. Relative to the uncoupled case, the latter effect has roughly twice the amplitude of the former, since the fragmentation process occurs by splitting an originally mixed CDM object into two separate halos dominated by the two different CDM particle types, respectively, and with roughly half the mass of the parent structure, such that for any intermediate mass halo that  disappears two new half-mass objects will populate the low-mass end of the halo mass function. Our results also showed for the first time how the
halo fragmentation process evolves in a hierarchical fashion, with halos of small and intermediate mass starting to fragment first, followed by progressively larger mass halos at lower redshifts. This is clearly shown by the shift of the transition between low-mass enhancement and high-mass suppression of the halo abundance to progressively larger masses for decreasing redshifts. Clearly, the strong enhancement of low-mass CDM halos in Multi-coupled Dark Energy cosmologies might be inconsistent with the observed abundance of satellite galaxies at small scales, thereby allowing to put constraints on the model. It is however important to point out here that it is not obvious how the baryonic and stellar components might evolve during the fragmentation process of their host CDM halo, thereby making it difficult to naively apply the standard abundance matching between visible galaxies and CDM halos also in the context of Multi-coupeld Dark Energy models. Dedicated radiative hydrodynamical simulations with cooling and star formation would be required to assess this point, and will be pursued in future works.

Another very interesting effect of Multi-coupled Dark Energy models on the abundance of collapsed structures concerns the mass range of galaxy clusters for large values of the coupling. In particular, for a coupling of order unity our simulations have shown a very significant suppression of the abundance of cluster sized CDM halos at low redshifts. Such feature might alleviate the present tension between the cosmologically inferred value of $\sigma _{8}$ and its best-fit value based on cluster counts, as reported also by the recent results of the Planck satellite mission. Clearly, a coupling value of $|\beta | = 1$ for Multi-coupled Dark Energy models might be disfavoured by other observable features such as e.g. the strongly enhanced abundance of low-mass halos. However, it is interesting to notice that our cosmological scenario might determine a suppression of the abundance of massive clusters without changing the large-scale normalisation of linear density perturbations, as such effect cannot be obtained in most of the other available Dark Energy or Modified Gravity models.
\begin{center}
{\em -- Structural properties of CDM halos --}
\end{center}
By matching individual structures in the different simulations, we have compared the structural properties of halos at different masses, taking the spherically averaged mass distribution around the most bound particle of each halo. Even in this case, our results have shown how for coupling values below the gravitational coupling threshold the overall impact of the DE-CDM interaction on the structural properties of halos is very mild, while it becomes significant for larger values of the coupling. In particular, we have investigated the total CDM density profile of halos, showing that for sub-gravitational couplings the density profile is almost unaffected and consistent with a standard NFW universal shape as for an uncoupled cosmology. On the contrary, larger values of the coupling determine a significant suppression of the overdensity in the inner part of the halos, and a corresponding increase of the density in the outer regions. Such effect is related to the outflow of the sub-dominant CDM species during the halo fragmentation process. We also found that, consistently with the previous results on the onset of the halo fragmentation process, halos in sub-gravitational coupling models are still composed by a mixture of the two distinct CDM particle species, while above the gravitational coupling threshold halos are dominated by a single CDM type at $z=0$. The resulting total gravitational potential (i.e. the potential arising by the superposition of the standard Newtonian potential and of the attractive and repulsive potentials associated to the scalar fifth-forces) has a non-trivial shape around CDM halos which depends on the relative distribution of the two CDM species within each structure. This effect gives rise to a species-dependent and (consequently) space-dependent screening mechanism of the scalar fifth-force even at the nonlinear and highly nonlinear level. Differently from other screening mechanisms associated to various classes of modified gravity theories, however, the screening mechanism of Multi-coupled Dark Energy cosmologies is a transient phenomenon, as it is based on the balance between attractive and repulsive corrections to standard gravity, and is therefore no longer effective as soon as mixed-CDM halos fragment into single-species objects. Nonetheless, for sub-gravitational coupling models such screening mechanism can hold until the present time, and provides a very effective suppression of the scalar fifth-forces.\\

To conclude, we have presented in this paper the first high-resolution N-body simulations of the Multi-coupled Dark Energy scenario, and investigated how such cosmological models affect the statistical and structural properties of collapsed objects forming from primordial density perturbations. Most remarkably, we have shown how the formation and evolution of cosmic structures is practically indistinguishable from a $\Lambda $CDM cosmology for coupling values below the gravitational coupling threshold $|\beta |=\sqrt{3}/2$. When such barrier is reached and overcome, our results have shown a series of very significant effects on structure formation, ranging from the fragmentation of previously bound halos into smaller objects, to the explicit violation of the weak equivalence principle, to the distortion of the universal shape of the halo mass function, to the modification of the density and gravitational potential profiles of halos that tend to become less overdense in their core. Finally, we have shown how Multi-coupled Dark Energy models feature a new type of nonlinear screening mechanism of the scalar fifth-forces associated with the coupling to Dark Energy, and how such screening mechanism is an unstable transient phenomenon that breaks down at the onset of halo fragmentation.

\section*{Acknowledgements}
I am deeply thankful to Luca Amendola for useful discussions on the Multi-coupled Dark Energy scenario, and to Frazer Pearce for useful suggestions on the generation of initial conditions for the N-body runs presented in this paper. The present work has been carried out under the  Marie Curie Intra European Fellowship
``SIDUN"  within the 7th Framework  Programme of the European Commission, and partly supported also by the INAF - Osservatorio Astronomico di Bologna and by the INFN-PD51 research project. All the numerical simulations presented here have been performed at the RZG supercomputing centre in Garching.

\footnotesize
\bibliographystyle{elsarticle-harv}
\bibliography{baldi_bibliography.bib}







\end{document}